\documentclass[aps,pre,showpacs,twocolumn]{revtex4-1}

\usepackage{amsmath,amssymb,amsfonts,hyperref}
\usepackage{graphicx,epsf}
\usepackage{xspace}
\usepackage{textcomp}

\newif\ifhyper
\hypertrue
\ifhyper
\hypersetup{
  citecolor = {green},
  urlcolor = {blue} 
} 

\newcommand{\be}{\begin{equation}}
\newcommand{\ee}{\end{equation}}

\newcommand{\p}{\partial} 
\newcommand{\vx}{\vec{x}}

\newcommand{\bq}{{\bf q}}
\newcommand{\bp}{{\bf p}}
\newcommand{\bx}{{\bf x}}
\newcommand{\vq}{{\vec{q}}}

\newcommand{\td}{\text{\tiny $D$}}

\newcommand{\tih}{\tilde{h}}
\newcommand{\tj}{\tilde{j}}
\newcommand{\tom}{\hat{\omega}}
\newcommand{\tnu}{\hat{\varpi}}
\newcommand{\tp}{\hat{p}}
\newcommand{\tq}{\hat{q}}
\newcommand{\tf}{\hat{f}}
\newcommand{\tzeta}{\hat{\zeta}}
\newcommand{\fhat}{\tilde{f}}
\newcommand{\frond}{\mathring{f}}
\newcommand{\fy}{f}
\newcommand{\Fhat}{\tilde{F}}
\newcommand{\Frond}{\mathring{F}}
\newcommand{\Fy}{F}
\newcommand{\Ftilde}{\bar{F}}

\newcommand{\etan}{\eta_{\nu}}
\newcommand{\etad}{\eta_{\td}}

\newcommand{\anz}{{ansatz}\xspace}

\newcommand{\nequ}{nonequilibrium\xspace}
\newcommand{\npt}{nonperturbative\xspace}
\newcommand{\nprg}{NPRG\xspace}
\begin{document}

\title{nonperturbative renormalization group for the Kardar-Parisi-Zhang equation: 
general framework and first applications}

\author{L\'eonie Canet$^1$, Hugues Chat\'e$^2$, Bertrand Delamotte$^3$ and  Nicol\'as Wschebor$^4$}
\affiliation{$^1$  Laboratoire de Physique et Mod\'elisation des Milieux Condens\'es, CNRS UMR 5493, Universit\'e Joseph Fourier Grenoble I, BP166,  38042 Grenoble Cedex, France\\$^2$ CEA, Service de Physique de l'\'Etat Condens\'e, Centre d'\'Etudes de  Saclay, 91191 Gif-sur-Yvette, France\\ $^3$ Laboratoire de Physique Th\'eorique de la Mati\`ere Condens\'ee, CNRS UMR 7600, Universit\'e Pierre et Marie Curie, 75252 Paris Cedex 05, France \\$^4$Instituto de F\'isica, Facultad de Ingenier\'ia, Universitad de la Rep\'ublica, J.H.y Reissig 565, 11000 Montevideo, Uruguay}

\begin{abstract}
We present an analytical method, rooted in the  \npt renormalization group, that allows one to  calculate  the critical exponents  and  the correlation and response functions of  the Kardar-Parisi-Zhang (KPZ) growth equation in all its different regimes, including the  strong-coupling one. We analyze the symmetries of the KPZ problem and derive an approximation scheme that satisfies the linearly realized ones. We  implement this scheme  at the minimal order in the response field, and show that it yields a complete, qualitatively correct phase diagram in all dimensions, with reasonable values for the  critical  exponents in physical dimensions. We also compute in one dimension the full (momentum and frequency dependent) correlation function, and  the associated universal scaling functions. We find an excellent quantitative agreement with the exact results from  Pr\"ahofer and Spohn \cite{spohn04}. %In particular, we obtain for the universal amplitude ratio $g_0\simeq 1.19(1)$, to be compared with the exact value $g_0=1.1504...$ (the Baik-Rain constant \cite{baik00}). 
We emphasize that all these results, which can be systematically improved, are obtained with sole input the bare action and its symmetries, without further assumptions on the existence of scaling or on the form of the scaling function.
\end{abstract}
%\date{\today}
\maketitle

\section{Introduction}

In their seminal work \cite{kardar86}, Kardar, Parisi and Zhang proposed a stochastic
continuum equation to describe surface growth through ballistic deposition which reads
\begin{equation}
\frac{\p h(t,\vx)}{\p t} = \nu\,\nabla^2 h(t,\vx) \, + 
\,\frac{\lambda}{2}\,\big(\nabla h(t,\vx)\big)^2 \,+\,\eta(t,\vx)
\label{eqkpz}
\end{equation}
where $\eta$ is an uncorrelated white noise of strength $D$, 
$\big\langle \eta(t,\vx)\eta(t',\vx')\big\rangle = 2\,D\,\delta^d(\vx-\vx')\,\delta(t-t')$,
 which models randomness in deposition. The term $\nu\,\nabla^2 h(t,\vx)$ provides the interface
 with a smoothening mechanism. The insightful feature of Eq.\ (\ref{eqkpz}) is to take into account the 
nonlinearity of the growth velocity through the inclusion of the term $\lambda(\nabla h(t,\vx))^2$
 which plays an essential role in the large scale properties of the height profile $h(t,\vx)$.

The KPZ equation (\ref{eqkpz}) is maybe the simplest nonlinear Langevin equation showing non-trivial behavior 
\cite{halpin95}, and as a consequence 
it arises in connection with an extremely large class of nonequilibrium or
disordered systems \cite{halpin95} such as randomly stirred fluid (Burgers equation) \cite{forster77},
  directed polymers in random media \cite{kardar87},  dissipative transport \cite{beijeren85,janssen86} or 
magnetic flux lines in superconductors \cite{hwa92}.  The KPZ equation has thus emerged as one of the 
fundamental theoretical models to investigate universality classes in \nequ scaling phenomena and phase transitions \cite{halpin95}. 
It is only recently, though, that a definitively-convincing experimental realization
has been brought out for a one-dimensional interface, confirming detailed theoretical predictions \cite{kaz10}.

The profile of the stationary interface is usually  characterized by the two-point correlation function 
\be 
C(t,|\vx|) \equiv \langle [h(t,\vx) - h(0,0)-t\langle \p_t h \rangle]^2\rangle\label{defC}
\ee
 and, in particular, by  its large scale  properties. The KPZ growth leads to generic scaling. At long 
time $\tau$ and large distance $L$,   $C$  assumes the scaling form 
$C(\tau,L) \propto \tau^{2\chi/z}\,g(L^{2\chi}\tau^{-2\chi/z})$  without fine-tuning any parameter of the model. 
The scaling function $g$ is universal and has the asymptotics  $g(y)\to$ const. as $y\to 0$ and $g(y)\sim |y|$ as $y\to \infty$. 
%, so that $C$ grows with time like $\tau^{\chi/z}$ at early stages of the growth until it saturates to $L^\chi$ 
% when $t\sim L^z$. 
$\chi$ and $z$ are the universal roughness and dynamical exponents respectively. 
 In fact, these two exponents are not independent since the Galilean symmetry \cite{forster77} --- the 
invariance of Eq.\ (\ref{eqkpz}) under an infinitesimal tilt of the interface --- enforces the scaling relation $z+\chi = 2$ as long as $\lambda\neq 0$.

The KPZ equation encompasses two distinct scenarios depending on the 
dimension of the interface.
Above two dimensions, there exist two different regimes 
separated by a critical value  $\lambda_{\rm c}$ of the nonlinear 
coefficient \cite{kardar86,forster77}. In the weak-coupling regime 
($\lambda < \lambda_c$),  the interface remains smooth, its properties 
are determined by the $\lambda= 0$ (Gaussian) fixed point ---  corresponding to the linear  
Edwards-Wilkinson equation \cite{halpin95} 
 --- with exponents $\chi = (2-d)/d$ 
and $z=2$. In the strong-coupling regime ($\lambda > \lambda_{\rm c}$), 
the nonlinearity becomes relevant and the interface roughens. 
In this regime, the exponents are not known exactly,
and some important issues are still controversial, 
such as the existence of an upper critical dimension
(for a recent discussion, see {\it e.g.\ }\cite{katzav02}). 
This unsatisfactory situation has persisted up to very recently \cite{canet10} because 
the strong coupling phase 
of the KPZ equation has remained out of reach of controlled analytical approaches. 
In particular, standard perturbation expansions \cite{kardar86,medina89,frey94} 
 have been proved to fail {\it at all order} to find a 
strong coupling fixed point \cite{wiese98}. Some nonperturbative approaches have  
been devised, such as the mode-coupling (MC)  approximation 
\cite{beijeren85,frey96,bouchaud93,doherty94,colaiori01}, or the weak noise 
scheme \cite{fogedby05}, but they are difficult to improve 
in practice.  (Some specific comments regarding the MC  approximation are put forward throughout this paper).
Let us finally mention the self-consistent expansion,
which is performed on the Fokker-Planck equation and can be systematically improved  \cite{SCE1,SCE2,katzav04}. It has yielded important results  for the KPZ equation in arbitrary dimensions, where it predicts an upper critical dimension \cite{katzav02}. Note that
 it has also been applied to different extensions of the KPZ model, such as  the non local KPZ equation \cite{SCE1} or the KPZ equation with colored noise \cite{SCE2}.

In two dimensions and below, the situation is different, both from a physical 
and a theoretical point of view. Physically,   the interface always roughens, there is no transition.  
On the theoretical side, exact results are available in one dimension 
\cite{baik00,spohn04,sasamoto05,spohn11,calabrese11}.
 The critical exponents  $\chi=1/2$ and $z=3/2$ have been known for long since they are fixed 
 by the existence of an incidental fluctuation-dissipation theorem in this dimension
\cite{kardar86}, but the scaling function $g(y)$ (and other universal properties)
 have  been computed exactly only very recently \cite{baik00,spohn04,spohn11,calabrese11}.
 Note that there had been earlier attempts to determine the scaling function, 
in the framework of the mode-coupling  approximation
 \cite{hwa91,frey96,colaiori02}.  
 In Ref. \cite{colaiori02}, a refined  ansatz for the scaling function in one dimension is 
devised to solve self-consistently the MC equations and it turns out 
that the result compares quite accurately with the exact solution. 
The one-dimensional scaling function has also been obtained   within the self-consistent expansion in one-dimension \cite{katzav04} and coincides in many respects with the MC result (see Section \ref{NLO}).
 Regarding the  MC theory, it is an `adhoc' approximation which consists of resumming one-loop 
diagrams while discarding vertex corrections.
% is difficult to control since it does not rely on a small parameter expansion and cannot in practice be systematically improved by calculating higher orders. 
The quality of the results is all the more surprising that it was shown 
that the contribution of the neglected terms are of the same order as 
those kept \cite{frey96}. The main drawbacks of the MC approach are that  
it strongly relies on the quality of the `educated guess' for the ansatz
 -- only available in one dimension up to now -- and it cannot in practice be 
systematically improved by calculating higher orders.

Recently, we have proposed an analytical approach to the KPZ equation based on  
\npt renormalization group (\nprg) techniques \cite{canet10}.
This early work has shown that the NPRG flow equations embedded a strong-coupling fixed point in all dimensions, and  it has  thus provided  a complete, qualitatively correct 
phase diagram for the first time within a RG approach, as well as reasonable values for the 
strong-coupling fixed point exponents in physical dimensions \cite{canet10}. 

In this paper we present a general and  systematic framework
 for applying NPRG methods to the KPZ problem, 
 strongly constrained by the symmetries of this model.
We derive general Ward identities and introduce a `covariantization' 
associated with the Galilean symmetry which, to the best of our knowledge, 
has never been reported before in the literature.
Within this formalism, successive orders of approximation are easily made explicit. % (although probably difficult to implement at high orders). 
In the present contribution, 
we  implement the minimal
order in the response field of this approximation scheme. We review the (simplified) first account of it
   presented in \cite{canet10} -- postponing its complete revisited version -- and derive 
new results related to the one-dimensional problem,
 in order to confront our approach to the available exact results. In particular,
we compute the correlation function, show that it takes a scaling form at long time and large distance and extract the associated scaling function.
  The NPRG results are  in even
better agreement with the exact results of Pr\"ahofer and Spohn \cite{spohn04} than those obtained under the MC or the self-consistent approximations. 

We stress the advantages of the NPRG formalism: 
(i) it is based on an exact flow equation, which, given a microscopic model, 
yields the macroscopic properties of the system. The only input is the bare action,  
that is,  no {\it a priori} knowledge, other than the microscopic model and its symmetries,  
is required to compute the physical observables. In particular, 
one does not have to assume scaling at long time and large distance 
 nor the form of the scaling function;
 (ii) the approximations, which are mandatory to solve the exact flow equation
can be systematically improved and implemented in any dimension and, in contrast to most other nonperturbative approaches, the calculation of higher orders is achievable and have been performed in practice in other systems \cite{berges02,canetpms}; 
(iii) beyond scaling, a wealth of quantities can be calculated (for example, 
we compute here the correction to scaling exponent $\omega$); (iv) it has already 
yielded  many  nontrivial and accurate results in systems at equilibrium
 such as frustrated magnets \cite{delamotte03}, the random bond and random field Ising 
model \cite{Tarjus2004},
membranes \cite{Essafi2011}, bosonic systems \cite{Dupuis2009},
but also in systems out-of-equilibrium, where one can mention
important advances in reaction-diffusion systems \cite{canet04}.\\
%The main drawbacks of the MC approach are hence circumvented using the NPRG.

As the NPRG is a field theoretical method, our starting point is the field theory associated with
 Eq.\ (\ref{eqkpz}), which follows from the standard procedure of Janssen-de Dominicis 
relying on the introduction of a response field  $\tih$ and sources $(j,\tj)$ \cite{janssen76}. This procedure allows one to explicitly carry out the integration over the Gaussian-distributed noise $\eta$ in Eq.\ (\ref{eqkpz}) upon  doubling  the number of fields (see {\it e.g.\  }\cite{canet11}).
The generating functional reads:
\begin{eqnarray}
{\cal Z}[j,\tj] \!\! &=& \!\!\!\int {\cal D}[h,i \tih]\, 
\exp\left(-{\cal S}[h,\tih] +  \int_{\bf x} j h+\tj\tih \right)\label{Z}\\
{\cal S}[h,\tih]  \!\! &=& \!\!\! \int_{\bf x}  \left\{ \tih\left(\p_t h -\nu \,\nabla^2 h - 
\frac{\lambda}{2}\,({\nabla} h)^2 \right) - D\, \tih^2  \right\}
\label{ftkpz}
\end{eqnarray} 
where ${\bf x}=(t,\vx)$.

The remainder of the paper is organized as follows. 
In Section \ref{NPRG}, we briefly review the \npt renormalization group formalism for out-of-equilibrium problems.
In Section \ref{SYM}, we analyze in detail the symmetries of the KPZ action (\ref{ftkpz}) 
and derive  Ward identities associated with the linearly realized ones. 
In Section \ref{ANZ}, we build an approximation scheme based on a covariantization
procedure rooted in the symmetries, and derive explicitly an  ansatz at the  minimal order in the response field. The  determination of the complete phase diagram and of critical exponents in all dimensions, 
 using a simplified version of this \anz,    is reported in Section \ref{LO}.
Section \ref{NLO} is then devoted to the calculation  (using the full  ansatz)
 of the scaling function and of some other universal quantities in one dimension, which are  compared with their exact counterparts.
Technical details, such as some of the Ward identities, a discussion of the validity of our approximation scheme, the computation of vertex functions or the procedures for the numerical 
integration of the flow equations are reported in Appendices A, B, C and D respectively. 

\section{The \npt renormalization group}
\label{NPRG}

The NPRG formalism relies on Wilson's RG idea, 
which consists of building  a sequence of scale-dependent  effective models   
such that fluctuations are smoothly averaged as the (momentum) scale $\kappa$ is lowered 
from the  microscopic scale $\Lambda$, where no fluctuations are yet included, to the macroscopic one $\kappa=0$ where  all fluctuations 
 are summed over \cite{berges02,delamotte07}.

For out-of-equilibrium problems, one formally proceeds as in equilibrium,
 but with the presence of additional response fields and with special care required to deal with  the consequences of It$\bar{\rm o}$'s discretization and with causality issues,
 as stressed in detail in \cite{canet11} --  from which conventions are taken throughout this paper.
For future use, we define the Fourier conventions used in \cite{canet11} and  throughout this work:
\begin{eqnarray}
\tilde{f}(\omega,\vec p) &=& \int d^d \vec x \,dt \;f(t,\vec x)\; e^{-i \vec p \cdot \vec x + i\omega t}\\
 f(t,\vec x) &=& \int \frac{d^d \vec p}{(2\pi)^d}\frac{d \omega}{2\pi}\;  \tilde{f}(\omega,\vec p)\; 
e^{i \vec p .\vec x - i\omega t}\\
&\equiv & \int_{\bp} \tilde{f}(\bp)\; e^{i \vec p .\vec x - i\omega t},
\end{eqnarray}
where $\bp=(\omega,\vec p)$.

To achieve the separation of fluctuation modes within the NPRG procedure, one  adds to the original action ${\cal S}$,
a momentum-dependent mass-like term:
\begin{equation}
\Delta {\cal S}_\kappa \!=\!\frac{1}{2}\! \int_{\bf q}\!  h_i(-{\bf q})\, 
[R_\kappa({\bf q})]_{ij}\, h_j({\bf q})\;\;\label{deltask}
\end{equation}
where the indices $i,j\in\{1,2\}$ label the field and response field respectively $h_1=h,h_2=\tih$, and 
summation over repeated indices is implicit. 
The matrix elements $[R_\kappa({\bf q})]_{ij}$ are proportional to a cutoff function
 $r(q^2/\kappa^2)$ (see Section \ref{cutcut}), with $q=\|\vec{q}\|$, which ensures the selection of 
fluctuation modes: $r(x)$ is required to vanish as $x\gtrsim 1 $ such that 
the fluctuation modes $h_i(q \gtrsim \kappa)$ are unaffected by 
$\Delta {\cal S}_\kappa$,
and to be large  when $x\lesssim 1 $ such that the other modes ($h_i(q\lesssim \kappa)$) are essentially frozen. Since $\Delta {\cal S}_\kappa$ must preserve 
the symmetries of the model, we postpone the discussion of the precise structure
of the matrix elements  $[R_\kappa({\bf q})]_{ij}$ to Section \ref{cutcut} after the analysis of these symmetries.

In presence of the mass term $\Delta {\cal S}_\kappa$, the generating functional (\ref{Z}) becomes scale dependent 
\begin{equation}
{\cal Z}_\kappa[j,\tj] \!\! = \!\!\!\int {\cal D}[h,i \tih]\, 
\exp\left(-{\cal S}-\Delta{\cal S}_\kappa+  \int_{\bf x} j h+\tj\tih \right)\label{Zk}
\end{equation} 
and the effective action $\Gamma_\kappa[\varphi,\tilde\varphi]$, where 
$\varphi_i=\langle h_i \rangle_{j,\tilde j}$ are 
the expectation values of the fields $h_i$ in the presence of the external sources $j$ and $\tilde{j}$,
 is given by the Legendre transform of  ${\cal W}_\kappa = \log {\cal Z}_\kappa$
(up to a term proportional to $R_\kappa$) \cite{berges02,canet11}:
\begin{equation}
\Gamma_\kappa[\varphi,\tilde\varphi] +\log {\cal Z}_\kappa[j,\tj] = 
\int\! j_i \varphi_i -\frac{1}{2} \int_{\bf q}\varphi_i\, [R_\kappa]_{ij}\, \varphi_{j}.
\label{legendre}
\end{equation}
 From $\Gamma_\kappa$, one can derive two-point correlation and response functions,
\be
[\,\Gamma_\kappa^{(2)}\,]_{i_1 i_2}({\bf x}_1,{\bf x}_2, \varphi,\tilde\varphi) = 
\frac{\delta^2 \Gamma_\kappa[\varphi,\tilde\varphi]}{\delta\varphi_{i_1}({\bf x}_1)\delta\varphi_{i_2}({\bf x}_2)}
\ee
and more generally $n$-point correlation functions that we write 
 here
in a $2\times2$ matrix form  (omitting the dependence on the running scale $\kappa$)
\begin{equation}
\Gamma_{i_3,...,i_n}^{(n)}({\bf x}_1,...,{\bf x}_n,\varphi,\tilde\varphi) =
\frac{\delta^{n-2} \Gamma^{(2)}({\bf x}_1,{\bf x}_2,\varphi,\tilde\varphi)}
{\delta\varphi_{i_3}({\bf x}_3)...\delta\varphi_{i_n}({\bf x}_n)} \;.
\end{equation}
The exact flow for $\Gamma_{\kappa}[\varphi,\tilde\varphi]$ is given by Wetterich's 
equation which reads (in Fourier space) \cite{berges02}:
\begin{equation}
\partial_\kappa \Gamma_\kappa \!=\! \frac{1}{2}\, {\rm Tr}\! \int_{\bf q}\! \partial_\kappa R_\kappa \cdot G_\kappa
\;\;{\rm with}\;\; G_\kappa\!=\!\left[\Gamma_\kappa^{(2)}+R_\kappa\right]^{-1}\!
\label{dkgam}
\end{equation}
 the full renormalized propagator of the theory.
When $\kappa$ flows from $\Lambda$ to 0, $\Gamma_\kappa$ interpolates 
between the microscopic model $\Gamma_{\kappa=\Lambda}={\cal S}$  and the full effective action $\Gamma_{\kappa=0}$
 that encompasses all the macroscopic properties of the system \cite{canet11}.
Differentiating Eq.\ (\ref{dkgam}) twice with respect to the fields
and evaluating it in a uniform and stationary field configuration (since the model is analyzed in its long time and large distance regime where it is translationally invariant in space and time)
one obtains the flow equation for the two-point functions:
\begin{eqnarray}
\partial_\kappa [\,\Gamma^{(2)}\,]_{ij}(\bp)\! &=& \! {\rm Tr}\! \int_{\bq} \partial_\kappa R(\bq) \cdot G(\bq) \cdot
\!\bigg(\!\!-\!\frac{1}{2}\, \Gamma^{(4)}_{ij}(\bp,-\bp,\bq) \nonumber\\
&& \hspace{-2.2cm} +\Gamma^{(3)}_{i}(\bp,\bq) \cdot G(\bp+\bq) \cdot
\Gamma^{(3)}_{j}(-\bp,\bp+\bq) \bigg) \cdot G(\bq)
\label{dkgam2}
\end{eqnarray}
where the $\kappa$ and background field dependencies have been omitted,
as well as the last  argument of the $\Gamma^{(n)}$ which is determined by
frequency and momentum conservation \cite{canet11}.\\

Solving Eq.\ (\ref{dkgam}) (or Eq.\ (\ref{dkgam2})) is in principle equivalent 
to solving the model. In practice this resolution cannot be performed exactly since (\ref{dkgam})  
is a nonlinear integral partial differential functional equation. 
Hence one has to devise an approximation scheme.
The main constraint on this approximation scheme is to preserve the symmetries
of the problem. We thus now revisit the symmetries of the KPZ action.

\section{Symmetries of the KPZ action}
\label{SYM}

The KPZ action (\ref{ftkpz}) possesses well-known symmetries, in addition to
translation and rotation invariances:
i) the Galilean symmetry and ii) the $h$-shift symmetry, which can be expressed as 
the invariance  of the action (\ref{ftkpz}) under the following transformations:
\begin{eqnarray}
\text{i)} && \left\{
\begin{array}{l}
h'(t,\vec x)=\vec x \cdot \vec v + h(t,\vec x+ \lambda \vec v t)\\
\tilde h'(t,\vec x)=\tilde h(t,\vec x+ \lambda \vec v t). \label{gal}
\end{array}
\right.\\
\text{ii)} && \;\;\;\; h'(t,\vec x)=h(t,\vec x)+c \label{shift}
\end{eqnarray}
where $\vec v$ and $c$ are arbitrary constant quantities. 

In one dimension, the KPZ equation also satisfies a fluctuation-dissipation 
theorem that fixes the exponents exactly.  This property roots in a 
time-reversal symmetry of the action which, as  shown in \cite{canet05},  can be encoded in the transformation 
\begin{equation}
\text{iii)} \left\{
\begin{array}{l}
h'(t,\vec x)=- h(-t,\vec x) \\
\tilde h'(t,\vec x)=\tilde h(-t,\vec x) +\frac{\nu}{D} \nabla^2 h(-t,\vec x).
\end{array}
\right. \label{fdt}
\end{equation}
One can check that the invariance of the action  (\ref{ftkpz}) under the transformation (\ref{fdt})
 requires that the contribution $\int \nabla^2 h (\nabla h)^2$ vanishes, which 
 is `incidentally' true only in $d=1$. The time-reversal symmetry thus only holds in this dimension.
This set of  symmetries  entails Ward identities for the $n$-point vertex functions, as long as the 
 mass term $\Delta{\cal S}_\kappa$ is appropriately chosen (see Section \ref{cutcut}).

In fact, there exist even stronger symmetries of the KPZ action that, 
to our knowledge, were only  pointed out in \cite{lebedev94}.
They consist of `gauging' in time the transformations i) and ii) in the 
following way:
\begin{eqnarray}
\text{i')} && \left\{
\begin{array}{l}
h'(t,\vec x)=\vec x \cdot \p_t \vec v(t) + h(t,\vec x+ \lambda \vec v(t))\label{galgauged}\\
\tilde h'(t,\vec x)=\tilde h(t,\vec x+ \lambda \vec v(t))
\end{array}
\right.\\
\text{ii')} && \;\;\;\; h'(t,\vec x)=h(t,\vec x)+c(t) \label{time}
\end{eqnarray}
which we will refer to as Galilean-gauged and shift-gauged symmetries respectively. 
In these gauged versions, $\vec v(t)$ and $c(t)$ are arbitrary infinitesimal functions of time.
Note that the action (\ref{ftkpz}) is not strictly invariant under the 
transformations (\ref{galgauged}) and (\ref{time}) but the corresponding 
variations of the action are linear in the fields, and this behavior also yields 
useful Ward identities, which we derive in the following.
These stronger forms of the symmetries will be thoroughly exploited.

Finally, an additional $Z_2$ symmetry, which is manifest on the Cole-Hopf version of the theory,  is  nonlinearly realized in the  KPZ action (\ref{ftkpz}).
 The Cole-Hopf field transformation writes as follows:
\begin{equation}
\left\{\begin{array}{l}
h({\bf x})=\frac{2 \nu}{\lambda} \log |w({\bf x})| \\
\tilde h({\bf x})=w({\bf x})\tilde w({\bf x}).
\end{array}
\right.
\end{equation}
In terms of $w$ and $\tilde w$ -- upon rescaling these fields and time --  the KPZ action becomes:
\begin{equation}
{\cal S}[w,\tilde{w}] = \int_{\bf x}  \tilde{w}(\p_t w -\nabla^2 w) - \frac 1 4 g_b\,( w\tilde{w})^2  
\label{cole-hopf}
\end{equation} 
with the bare coupling constant
\begin{equation}
\label{def-g}
g_b=\frac{\lambda^2 D}{\nu^3}.
\end{equation} 
This action is invariant under the simple $Z_2$ transformation (iv) $w(t,\vec{x})\to \tilde{w}(-t,\vec{x}) $,
 $\tilde{w}(t,\vec{x})\to {w}(-t,\vec{x}) $.
However, in terms  of  the original fields $h$ and $\tih$, this transformation  becomes
\begin{equation}
\text{iv)} \left\{
\begin{array}{l}
h'(t,\vec x)=- h(-t,\vec x)+\frac{2 \nu}{\lambda} \log |\tilde h (-t,\vec x)| \\
\tilde h'(t,\vec x)=\tilde h(-t,\vec x).
\end{array}
\right.
\label{z2}
\end{equation}
Its highly nonlinear form  renders complicated the 
 induced Ward identities among vertex functions. Consequently,   they are not given here as 
 they will  not be  exploited in the following.
 
\subsection{Shift-gauged symmetry}

The Ward identity associated with the shift-gauged symmetry  can be derived by 
 performing in the functional integral (\ref{Zk})
the change of variables corresponding to the transformation (\ref{time}).
As this operation must leave  the value of the integral unchanged, one obtains that
\begin{equation}
\int_{\bx} \Big\{ c(t) j({\bf x})- \langle \tilde h({\bf x})\rangle_{j,\tilde j} \partial_t c(t)\Big\}
 -\langle \Delta S_\kappa[c(t),\tilde h({\bf x})]\rangle_{j,\tilde j}
\end{equation}
must vanish.  Here,  $[R_\kappa({\bf q})]_{11}$ has been set to zero (as in Eq.\ (\ref{Rk}))  for simplicity and for causality issues (see Section \ref{cutcut}). 
As the mass term is quadratic in fields, 
$\Delta S_\kappa[c(t),\tilde h({\bf x})]$ is linear in $\tilde h$, and thus 
$$\langle \Delta S_\kappa[c(t),\tilde h({\bf x})]\rangle_{j,\tilde j}=\Delta S_\kappa[c(t),\langle\tilde h({\bf x})\rangle=
\tilde\varphi({\bf x})].$$
The expression (\ref{legendre}) of the Legendre transform then implies
\begin{equation}
\int_{\bx} \left\{ c(t) \frac{\delta \Gamma_\kappa[\varphi,\tilde\varphi]}{\delta \varphi({\bf x})}
- \tilde \varphi({\bf x}) \partial_t c(t)\right\}=0.
\end{equation}
After integrating by parts the second term, we conclude that the functional 
\begin{equation}
\Gamma_\kappa[\varphi,\tilde \varphi]-\int_{\bx} \;\;\tilde \varphi ({\bf x}) \partial_t \varphi ({\bf x})
\end{equation}
is invariant under the transformation  (\ref{time}). In other words, the only non-invariant term
 $\int \tilde \varphi \partial_t \varphi$  of the bare action is not renormalized and the rest 
of the action is  shift-gauged symmetric.

Let us express this property on the $n$-point vertex functions, that we denote from now on  as 
$$\Gamma_\kappa^{(l,m)}({\bf x}_1,\dots,{\bf x}_{l+m})$$
which stands for  the $\Gamma_\kappa^{(n=l+m)}$ vertex involving $l$ (respectively  $m$)
legs -- derivatives of $\Gamma_\kappa$ with respect to $\varphi$ (respectively  $\tilde\varphi$) -- with the $l$ first frequencies and 
momenta referring to the $\varphi$ fields and the $m$ last to the $\tilde\varphi$ fields.
At this stage, it is convenient to work in Fourier space.  
The $n$-point vertex function are defined in Fourier space by
\begin{widetext}
\begin{equation}
(2 \pi)^{d+1}\delta^{d+1}\big(\sum_i {\bf p}_i\big)  \Gamma_\kappa^{(l,m)}({\bf p}_1,\dots,{\bf p}_{l+m-1}) =  \int_{\bx_1\cdots \bx_{l+m}}\Gamma_\kappa^{(l,m)}({\bf x}_1,\dots,{\bf x}_{l+m})e^{i\sum_i (\vec x_i \cdot \vec{p}_i- t_i \omega_i)}
\end{equation}
\end{widetext}
where again, the last frequency and momentum, fixed by translational invariance in  time and space, are implicit.
The shift-gauged symmetry entails that the $n$-point vertex functions in Fourier space satisfy the following property:
\begin{equation}
\label{wardtemp}
\Gamma_\kappa^{(m,n)}(\omega_1,\vec p_1=0,\dots,{\bf p}_{m+n-1})= i\omega_1 \delta_{m1}\delta_{n1}
\end{equation}
which means that, apart from the contribution of  $\int \tilde \varphi \partial_t \varphi$  to 
 $\Gamma_\kappa^{(1,1)}$, the vertices vanish upon setting  the momentum of one  of the  $\varphi$ to zero.
 This is related to the fact that the field $\varphi$ only appears in $\Gamma_\kappa$  with gradients.
In particular, it roots the non-renormalization of the kinetic term $\tilde \varphi \partial_t \varphi$ of the KPZ action, which is well established in perturbation theory (see {\it e.g.\ }\cite{frey96}).

\subsection{Galilean symmetry}

\subsubsection{Global Galilean symmetry}

Let us first review the Ward identities associated with the standard Galilean symmetry.
As for the shift-gauged symmetry, one can prove that, as the transformation (\ref{gal}) encoding  the Galilean symmetry of the bare action is affine, and provided the mass term $\Delta {\cal S}_\kappa$ is chosen Galilean-invariant (as in the form (\ref{Rk}) -- see Section \ref{cutcut}), $\Gamma_\kappa$ also possesses this symmetry.
The corresponding Ward identity is:
\begin{multline}
\label{WardGalilee}
\int_{\bx} \left\{ \left(\vec x \cdot \vec v + \lambda t \vec v \cdot \nabla
\varphi({\bf x})\right)\frac{\delta \Gamma_\kappa}{\delta \varphi({\bf x})}\right. \\
 \left.+\lambda t \vec v \cdot \nabla \tilde \varphi ({\bf x})
\frac{\delta \Gamma_\kappa}{\delta \tilde \varphi({\bf x})}\right\} =0,
\end{multline} 
that is $\Gamma_\kappa$ is invariant under the same Galilean transformation (with the same parameter)
 as the bare action.
One can then proceed to the derivation of the Ward identities for the vertex functions, by taking functional derivatives of Eq.\ (\ref{WardGalilee}) with respect to
 the fields and then evaluating them for uniform and static fields.
For instance, one gets  the standard identity for the three-point function:
\begin{equation}
i\frac{\partial}{\partial \vec p}
\Gamma_\kappa^{(2,1)}(\omega=0,\vec p=\vec 0;\omega_1,\vec p_1)
= \lambda \vec p_1 \frac{\partial}{\partial \omega_1} \Gamma_\kappa^{(1,1)}(\omega_1,\vec p_1).
\label{wardgal21}
\end{equation}
The  Ward identity for a generic $n$-point function can be derived and one obtains:
\begin{widetext}
\begin{multline}
\label{galgen}
i\frac{\partial}{\partial \vec p}
\Gamma_\kappa^{(m+1,n)}(\omega=0,\vec p=\vec 0;{\bf p}_1;\dots;{\bf p}_{m+n-1})
\\
= \lambda \left(\vec p_1 \frac{\partial}{\partial \omega_1}
+\dots+\vec p_{m+n-1} \frac{\partial}{\partial \omega_{m+n-1}} \right)
\Gamma_\kappa^{(m,n)}({\bf p}_1;\dots;{\bf p}_{m+n-1}).
\end{multline}
\end{widetext}

\subsubsection{Galilean-gauged symmetry}

Let us now come to the gauged form (\ref{galgauged}) of the Galilean symmetry.
As for the shift-gauged symmetry, the variation of the action
 under this transformation is linear in the fields and consequently, it entails a Ward identity which reads
\begin{multline}
\label{WardGalileejaugee}
\int d^d\vec x \left\{ \lambda \nabla \varphi({\bf x})\frac{\delta \Gamma_\kappa}{\delta \varphi({\bf x})}
-\vec x \partial_t \frac{\delta \Gamma_\kappa}{\delta \varphi({\bf x})}\right.\\
\left.+\lambda \nabla \tilde \varphi ({\bf x})
\frac{\delta \Gamma_\kappa}{\delta \tilde \varphi({\bf x})}
-\vec x \partial_t^2 \tilde \varphi({\bf x})\right\}=0.
\end{multline}
 From this functional identity, one can again deduce identities for the vertex functions.
They bare similar expressions as those for the global Galilean symmetry, but with a stronger content.
For instance, the identity for the three-point function becomes:
\begin{multline}
\label{Gal21gauged}
i\omega \frac{\partial}{\partial \vec p}
\Gamma_\kappa^{(2,1)}(\omega,\vec p=\vec 0;\omega_1,\vec p_1)
= \lambda \vec p_1 \\
\times \left(\Gamma_\kappa^{(1,1)}(\omega+\omega_1,\vec p_1)-\Gamma_\kappa^{(1,1)}(\omega_1,\vec p_1)\right),
\end{multline}
which in the limit $\omega\to 0$ coincides with (\ref{wardgal21}).
The gauged identity (\ref{Gal21gauged}) is stronger as it constrains 
the whole frequency dependence and not only the zero-frequency sector.
One could derive similar identities for generic $n$-point functions, 
but we now stress, instead,  a more efficient way to exploit the 
Galilean symmetry, that will guide our construction of the approximation scheme.

\subsubsection{Covariant time derivatives}
\label{covariant}
The previous  Ward identities ensuing from the Galilean symmetry do not clearly reflect  the geometrical interpretation of this symmetry. In order to do so, one can analyze the Galilean invariance from another angle.
One can build quantities which are manifestly scalar under the Galilean
 transformation (\ref{gal}), upon introducing an adequate
 covariant time derivative.
 Let us define a function $f({\bf x})$ as a scalar under the Galilean symmetry
 if  its infinitesimal transform under (\ref{gal}) is given by
\begin{equation}
 \delta f({\bf x})= t\lambda \vec{v} \cdot \nabla f({\bf x}).
\end{equation}
With this definition, if $f$ is a scalar then $\int d^d \vec x f$ is invariant under the Galilean
transformation and can be used to build an action possessing the Galilean symmetry. 
It follows from this definition  that 
the response field $\tilde h$ is a scalar, but that the field $h$ is not 
 unless one takes two successive space derivatives $\nabla_i \nabla_j h$.
The gradient of a scalar remains a scalar, but not its time derivative since
\begin{equation}
 \delta (\partial_t f({\bf x}))= \lambda \vec{v} \cdot \big( t \nabla (\partial_t f({\bf x}))
+\nabla f({\bf x})\big).
\end{equation}
However, as in fluid mechanics, one can construct a covariant time derivative 
$$\tilde{D}_t \equiv \partial_t -\lambda \nabla h({\bf x})\cdot \nabla $$
 which conserves the scalar property, {\it i.e.\ } if $f$ is a scalar then so is $\tilde{D}_t f$.
 Note that the covariant time derivative of the field $h$ itself, that we denote $D_t$,  bares a special form with a 1/2 factor:
\begin{equation}
D_t h({\bf x})\equiv \partial_t h({\bf x})-\frac{\lambda}{2} (\nabla h({\bf x}))^2
\end{equation}
since $h$ is not a scalar on its own. 
These covariant derivatives will constitute the building blocks in the construction of 
our approximation scheme  (see Section \ref{ANZ}).

\subsection{Time-reversal symmetry in $d=1$}

In $d=1$, the action (\ref{ftkpz}) with a mass term of the form (\ref{Rk}) exhibits the additional time-reversal symmetry.
Note that this is a discrete symmetry ({\it i.e.\ } there is no infinitesimal transformation  corresponding to (\ref{fdt})).
However, as it is linear in the fields, one can show using the same procedure as previously
 that $\Gamma_\kappa$ also possesses this symmetry.
 That is, it verifies:
\begin{equation}
\label{WardTFD}
 \Gamma_\kappa[\varphi({\bf x}), \tilde \varphi({\bf x})]=\Gamma_\kappa[-\varphi(-t,\vec x),\tilde\varphi(-t,\vec x)
+\frac{\nu}{D}\nabla^2\varphi(-t,\vec x)].
\end{equation}
Again, one can derive Ward identities for the $n$-point vertex functions by taking derivatives of (\ref{WardTFD}) with respect to the fields and evaluating them at uniform and static field configurations.
For the two-point functions, this yields, in Fourier space:
\begin{eqnarray}
\label{WardTFD2}
&&\Gamma_\kappa^{(2,0)}(\omega,\vec p)
=\Gamma_\kappa^{(2,0)}(-\omega,\vec p)+\frac{\nu}{D}p^2\Gamma_\kappa^{(1,1)}(-\omega,\vec p)
\nonumber\\
&&+\frac{\nu}{D}p^2\Gamma_\kappa^{(1,1)}(\omega,-\vec p)
+\left(\frac{\nu}{D}\right)^2p^4\Gamma_\kappa^{(0,2)}(-\omega,\vec p),
\end{eqnarray}
which, given that $\Gamma_\kappa^{(2,0)}(\omega,\vec p)=\Gamma_\kappa^{(2,0)}(-\omega,\vec p)$ and spatial parity reduces to:
\begin{equation}
2\mathrm{Re}\Gamma_\kappa^{(1,1)}({\bf p})
=-\frac{\nu}{D}p^2\Gamma_\kappa^{(0,2)}({\bf p}).
\end{equation}
For the three-point functions, one gets:
\begin{eqnarray}
2 \mathrm{Re}\Gamma_\kappa^{(1,2)}({\bf p}_1;{\bf p}_2)
&=&-\frac{\nu}{D}p_1^2\Gamma_\kappa^{(0,3)}({\bf p}_1;{\bf p}_2).
\nonumber\\
2 \mathrm{Im}\Gamma_\kappa^{(2,1)}({\bf p}_1;{\bf p}_2)
&=&-\frac{\nu}{D}p_2^2\mathrm{Im}\Gamma_\kappa^{(1,2)}({\bf p}_1;{\bf p}_2)
\nonumber\\
&&\hspace{-1cm}-\frac{\nu}{D}p_1^2\mathrm{Im}\Gamma_\kappa^{(1,2)}({\bf p}_2;{\bf p}_1). \label{FDT21}
\end{eqnarray}
Similar expressions can be derived for $\Gamma_\kappa^{(3,0)}$ and for the four-point functions, which are reported for completeness in Appendix A.

\subsection{cutoff function}
\label{cutcut}

The mass term $\Delta {\cal S}_\kappa$ defined by (\ref{deltask}) 
must be chosen such that the Ward identities are preserved all along the flow. 
This is not obvious {\it a priori}.
 In particular,  a quadratic mass term cannot be invariant under the shift-gauged
 symmetry. However, as explained above, if the variation of $\Delta {\cal S}_\kappa$
 under the transformation (\ref{time}) is linear in the field, the Ward identity 
 remains identical even in the presence of such a  term. 
Similar comments hold for the Galilean symmetry, which is preserved as 
long as the cutoff matrix $R_\kappa$
 does not depend on the frequency. We thus choose a frequency independent matrix
 $R_\kappa$ -- which also preserves causality properties \cite{canet11}.
 Moreover, the time-reversal symmetry 
 imposes in one dimension the following relation between the components of the cutoff matrix:
 $[R_\kappa(\vq)]_{12}= -\frac{\nu}{2D}q^2 \, [R_\kappa(\vq)]_{22}$,
which we will impose in all dimensions.

Finally, we set $[R_\kappa(\vq)]_{11}=0$ since it is not necessary to consider
a cutoff term proportional to $\varphi\varphi$ and since the presence of such a term
would spoil the property that $\Gamma_\kappa$ is proportional to $\tilde\varphi$, important for causality issues (see \cite{canet11}) and also in order to preserve the shift symmetry.
To summarize, the cutoff matrix is chosen with the following form:
\begin{equation}
R_\kappa(\vq) \!=\! r\left(\frac{q^2}{\kappa^2}\right)
\left(\!\! \begin{array}{cc}
0& {\nu_\kappa} q^2\\
{\nu_\kappa} q^2 & -2 D_\kappa 
\end{array}\!\!\right) \;,
\label{Rk}
\end{equation}
where the running coefficients $\nu_\kappa$ and $D_\kappa$ are introduced for  convenience \cite{canet10}.
Here we use $r(x)=\alpha/(\exp(x) -1)$ where $\alpha$ is a free parameter.

\section{Approximation scheme}
\label{ANZ}

Two routes have been followed in the past to build approximation schemes
to deal with exact NPRG flow equations such as Eqs.\ (\ref{dkgam}) and (\ref{dkgam2}).
Either one focuses on the long time and large distance properties, 
trying to describe as accurately as possible the zero-momentum 
and zero-frequency domain while approximating the other sectors.
Or the aim is to compute the momentum and frequency dependence of
two-point functions and then the approximation 
 concerns the three- and four-point functions.

The usual implementation of the first route consists of performing a 
`derivative expansion': $\Gamma_\kappa$ is expanded in powers of 
gradients and time derivatives of the fields
\cite{berges02,canet11}. This strategy has been undertaken to investigate the critical properties of many 
equilibrium \cite{berges02}, 
and also \nequ systems such as reaction-diffusion  processes \cite{canet04}, 
where it has yielded satisfactory results.
However, for the KPZ problem,  the signature nonlinear term involves a gradient,
and this seems to preclude the use of the derivative expansion  \cite{canet05}.

The second route is the one followed here. At equilibrium, one can compute 
the momentum dependence of the two-point function using the 
Blaizot-M\'endez-Wschebor (BMW) approximation scheme which has proved to yield 
very accurate results for $O(N)$ models \cite{bmw}. In the BMW framework,
the momentum dependence of  $\Gamma_\kappa^{(3)}$ and  $\Gamma_\kappa^{(4)}$ in
the flow equation for  $\Gamma_\kappa^{(2)}$ is truncated.
However, a direct implementation of the BMW scheme for the KPZ problem is  hindered
  by the symmetries which impose strict identities between
$\Gamma_\kappa^{(m,n)}$'s with different $m$ and at different momenta, see {\it e.g.\ }Eq.\ (\ref{galgen}), and render complicated the writing of a truncation. 

To overcome this difficulty, we propose, instead of designing directly truncations on the three- and four- point functions, to construct an  ansatz for 
$\Gamma_\kappa$ that manifestly preserves the shift-gauged and Galilean-gauged 
symmetries while allowing for arbitrary frequency and momentum dependencies
in  the two-point functions. 
This is easily achieved by combining together the basic Galilean scalars of the theory --
$\tilde{\varphi}$, $\nabla_i\nabla_j \varphi$ and $D_t\varphi$, and arbitrary powers of their gradients
and covariant time derivatives $\tilde{D}_t$ -- since, by construction,
all calculations involving  a functional of these quantities automatically
satisfy the Galilean symmetry (the other symmetries, but the nonlinearly realized one, 
being easily enforced {\it a posteriori}). 

On this basis,  various truncations may be performed to obtain an \anz for $\Gamma_\kappa$. 
One could, for example, keep complete dependencies in $\tilde{\varphi}$ and
$\nabla^2\varphi$, while treating only the zero-frequency sector.
This `super derivative expansion' option is left for future studies.
An alternative choice, pursued here with the ultimate aim of confronting the obtained two-point
  correlation  functions  in one dimension with the exact ones, is to truncate the field dependence
of $\Gamma_\kappa$ while preserving the complete 
 dependence in both momentum and frequency of the two-point functions.

 In the following, we hence proceed to a field truncation at the minimal order.  It appears that the truncations of $\Gamma_\kappa$ at a given order in either $\varphi$ or $\tilde\varphi$ 
are not equivalent because the dependence in $\varphi$, contrary to that in $\tilde{\varphi}$, 
is  constrained by the Galilean symmetry --  $\varphi$ enters in $\tilde{D}_t$.
The minimal order in fields amounts to keeping terms, (i) at most quadratic in $\tilde{\varphi}$, (ii)  linear in $\nabla^2\varphi$ or  in $D_t\varphi$, and (iii) 
 to combine them with  arbitrary powers of  Laplacian $\nabla^2$ and
  covariant time derivatives $\tilde{D}_t$. This choice allows for  an arbitrary dependence in 
 momentum and frequency of the two-point functions. Notice that the dependence in $\varphi$
is not restricted to be polynomial since arbitrary powers of this field
 are included through the covariant derivative $\tilde{D}_t$.
 To summarize, the  ansatz for $\Gamma_\kappa$
 considered in the following  reads:
\begin{widetext}
\begin{equation}
\Gamma_\kappa[\varphi,\tilde \varphi]= \int_{\bf x} 
\left\{ \tilde \varphi f_\kappa^\lambda(-\tilde D_t^2,-\nabla^2) D_t\varphi - 
\tilde \varphi f_\kappa^D(-\tilde D_t^2,-\nabla^2) \tilde \varphi 
 - \frac{\nu}{2D} \left[\nabla^2 \varphi f_\kappa^\nu(-\tilde D_t^2,-\nabla^2) \tilde \varphi + 
 \tilde \varphi f_\kappa^\nu(-\tilde D_t^2,-\nabla^2)  \nabla^2 \varphi\right] \right\}. 
\label{anznlo}
\end{equation}
\end{widetext}
Setting $f_\kappa^\lambda=1$ and $f_\kappa^\nu=f_\kappa^D=D$ in the previous \anz, one is left with the bare action (\ref{ftkpz}) -- written in terms of covariant time derivatives. Allowing non-constant (and scale-dependent) functions $f_\kappa^X$ ($X=D,\nu,\lambda$) of the Laplacian  and of covariant derivatives is then the way to include an arbitrary dependence in momentum and frequency for the two-point functions that preserves all the symmetries, at the lowest (second) order in $\tilde\varphi$.

The operator $\tilde D_t$ is squared in $f_\kappa^D$ since this function is real. For $f_\kappa^\nu$
 and $f_\kappa^\lambda$, to take the square of $\tilde D_t$ is a choice that ensures a 
natural separation between real and imaginary parts of $\Gamma_\kappa^{(1,1)}$, without loss 
of generality for the two-point functions.
Note that one needs to make precise the meaning of the expressions 
$f_\kappa^X(-\tilde D_t^2,-\nabla^2)$ ($X=D,\nu,\lambda$). Indeed, it is ambiguous as it stands since the $\tilde D_t$ and $\nabla$ operators do not commute.
We take the convention that all $\tilde D_t$'s are on the left of the $\nabla$'s.
We also assume that these functions can be expanded in series of their arguments:
\begin{equation}
\label{serieformelle}
 f_\kappa^X(-\tilde D_t^2,-\nabla^2)=\sum_{m,n=0}^\infty a_{mn} (-\tilde D_t^2)^m (-\nabla^2)^n.
\end{equation}

One can then consider the constraints stemming from the other symmetries.
The shift-gauged symmetry  imposes that $f_\kappa^\lambda(\omega^2,\vec p^{\,2}=0) \equiv 1$.
In dimension one, the time-reversal symmetry implies that the two functions 
$f_\kappa^D$ and $f_\kappa^\nu$
 become identical,  and also that $f_\kappa^\lambda(\omega^2,\vec p^{\,2})\equiv 1$.
In a generic dimension $d$, the \anz (\ref{anznlo})  thus consists of three independent 
running functions of $p^2$ and $\omega^2$, and  is reduced to a unique function of 
$p^2$ and $\omega^2$ in $d=1$.

To compute the flow equations of the functions $f_\kappa^X(-\tilde D_t^2,-\nabla^2)$, $X=\lambda, \nu$ or $D$, one needs the expressions of the $n$-point functions up to $n=4$. 
All the calculations are performed at vanishing  fields.
For the  two-point functions, they are straightforward and yield
\begin{eqnarray}
\Gamma_\kappa^{(2,0)}(\omega,\vec p) &=& 0 \nonumber\\
\Gamma_\kappa^{(1,1)}(\omega,\vec p) &=& i\omega f_\kappa^\lambda\left(\omega^2, \vec p\,^2\right) +\frac{\nu}{D} \vec p\,^2 f_\kappa^\nu(\omega^2,\vec p\,^2)  \label{anzgam2}\\
\Gamma_\kappa^{(0,2)}(\omega,\vec p) &=& -2  f_\kappa^D(\omega^2,\vec p\,^2).  \nonumber
\end{eqnarray}
In this case, there is no ambiguity arising  from  the ordering of the gradient and 
covariant time derivative operators.
The calculation and the explicit expressions of the three- and four-point functions, 
more lengthy, are detailed in Appendix C.
One can check that all the Ward identities derived in Section \ref{SYM} are satisfied by these functions.

We can be  {\it a priori} confident in our approximation
scheme, as it is close in spirit to the BMW approximation which can accurately capture the momentum dependence of two-point functions \cite{bmw}. A more detailed discussion on the validity of our approximation scheme can be found in  Appendix B.

\section{Simplified approximation}
\label{LO}
The  ansatz (\ref{anznlo}) in  generic dimensions $d$ can be further simplified   by restraining the 
form of the functions $f_\kappa^X(-\tilde D_t^2,-\nabla^2)$. 
The idea is to focus on a reliable description of the 
 zero-frequency and zero-momentum sector of the theory while circumventing the
 limitations of the derivative expansion which is problematic here.
 For this, one can neglect all $\tilde D_t^2$ dependence in these functions:
$f_\kappa^X(-\tilde D_t^2,-\nabla^2) \to f_\kappa^X(-\nabla^2)$ for $X=\lambda, \nu$ and $D$.

The  ansatz  proposed in \cite{canet10} -- which was derived before the formalization  of the Galilean-gauged symmetry in terms of covariant derivatives --  is actually endowed with an additional  simplification
since we imposed the one-dimensional constraint $f_\kappa^\lambda=1$ in {\it all} dimensions
(or stated otherwise we extended the condition $f_\kappa^\lambda(\vec p=0)=1$ holding for all $d$ to the whole $\vec p$ sector). It reads
\begin{eqnarray}
\Gamma_\kappa[\varphi,\tilde \varphi]&=& \int_{\bx} \left\{ \tilde \varphi({\bf x}) D_t\varphi({\bf x})
-\tilde \varphi({\bf x})f_\kappa^D(-\nabla^2) \tilde \varphi({\bf x})\right. \nonumber\\
&&-\left. \frac{\nu}{D}\tilde \varphi({\bf x})f_\kappa^\nu(-\nabla^2)  \nabla^2 \varphi({\bf x})
 \right\} \;.
\label{anzlo}
\end{eqnarray}
As a matter of fact, this \anz implies that all interaction vertices 
($\Gamma_\kappa^{(n)}$ with $n>2$) are reduced to their bare form. Among these, the only non-vanishing  one 
is $\Gamma_\kappa^{(2,1)}$:
\begin{equation}
 \Gamma_\kappa^{(2,1)}(\omega_1,\vec p_1,\omega_2,\vec p_2)=\lambda \vec p_1\cdot \vec p_2.
\label{anz21}
\end{equation}
With the \anz (\ref{anzlo}), the two-point functions in generic dimensions  simplify to
\begin{eqnarray}
&&\Gamma_\kappa^{(2,0)}(\omega,\vec p)
=0 \nonumber\\
&&\Gamma_\kappa^{(1,1)}(\omega,\vec p)
=i\omega+ \frac\nu D \vec p\,^2 f_\kappa^\nu(p^2) \label{anzgam2lo}\\
&&\Gamma_\kappa^{(0,2)}(\omega,\vec p)
=-2 f_\kappa^D(p^2).
\nonumber
\end{eqnarray}
With the aim of analyzing the fixed point structure, one introduces dimensionless and renormalized  functions
 $\tf_\kappa^\nu\equiv f_\kappa^\nu/\nu_\kappa$  (respectively  $\tf_\kappa^D \equiv f_\kappa^D/D_\kappa$), where  $\nu_\kappa$ (respectively  $D_\kappa$) are  running coefficients (identifying at the microscopic scale $\kappa=\Lambda$ with the bare parameters $\nu$ (respectively  $D$) of the action (\ref{ftkpz})) 
 that are related to anomalous dimensions  $\kappa\partial_\kappa\nu_\kappa=-\etan(\kappa)\nu_\kappa$ (respectively  $\kappa\partial_\kappa D_\kappa=-\etad(\kappa)D_\kappa$).   At a fixed point, these coefficients are expected to behave as power laws $\nu_\kappa\sim \kappa^{-\etan^*}$ and  $D_\kappa\sim \kappa^{-\etad^*}$ and 
 the scaling exponents are then expressed in terms of these anomalous dimensions  as $z=2-\etan^*$  and $\chi = (2-d+\etad^*-\etan^*)/2$. 

The flow equations for the functions $\tf_\kappa^\nu$ and $\tf_\kappa^D$  can be found in \cite{canet10}. The Galilean invariance  ensures that the flow of the dimensionless running coupling constant $\hat{g}_\kappa\equiv \kappa^{d-2}\lambda^2 D_\kappa/\nu_\kappa^3$ is reduced to its dimensional part
\begin{equation}
\partial_s \hat{g}_\kappa = \hat{g}_\kappa (d-2+3\etan(\kappa)-\etad(\kappa))\;
\label{eqg}
\end{equation} 
with $\partial_s=\kappa\partial_\kappa$, which thus enforces the identity $z+\chi=2$ at any non-Gaussian fixed point.
 The subleading exponent $\omega$, governing the corrections to scaling,
  is independent
of the leading critical exponents $\chi$ and $z$ and can be calculated from the flow behavior as $s \to -\infty$, using  for instance that
\begin{equation}
\hat g_\kappa \sim \hat g^*+ \hat g_1 \exp(\omega s)
\label{calomega}
\end{equation}
 in this limit.

\begin{figure}[tp]
\epsfxsize=8.cm
\epsfbox{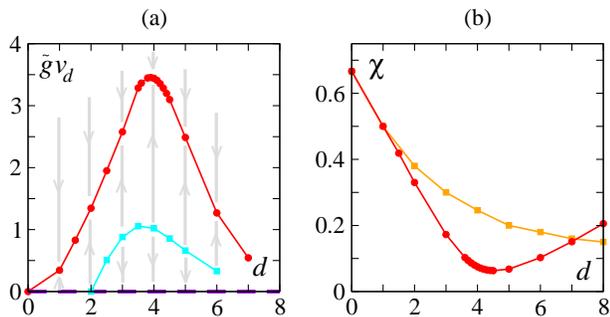}
\caption{(Color online) 
(a) Flow diagram stemming from  our (simplified) minimal order approximation
 in the $(d,\hat g v_d)$ plane (with $v_d^{-1}=2^{d+1}\pi^{d/2}\Gamma[\frac{d}{2}]$ a normalization factor  related to the integration volume). 
Red circles: renormalized value $\hat{g}_{\rm SC}^*$ at $F_{\rm SC}$.
Dashed purple line: Gaussian  fixed point $F_{\rm EW}$.
Cyan squares: bare value $\hat{g}_{\rm c}$,  which
separates the basins of attraction of $F_{\rm SC}$ and  $F_{\rm EW}$. Gray arrows
symbolize flow lines.
(b) Variation with $d$ of $\chi=2-z$ for $F_{\rm SC}$
(red circles: our results; orange squares: 
numerical values from \cite{tang92,castellano98}. See also Table~\ref{tab1}.)}
\label{fig1}
\end{figure}

Despite its simplicity, this \anz yields the correct phase diagram in all dimensions (Fig.~\ref{fig1}(a)) -- including the strong coupling fixed point,
and   reasonable 
exponent values in physical dimensions (Fig.~\ref{fig1}(b)) \cite{canet10}.
More precisely, in all dimensions studied ({\it i.e.\ } up to $d=8$), there exists
besides the Edwards-Wilkinson fixed point $F_{\rm EW}$,
a fully-attractive non-trivial strong-coupling fixed point $F_{\rm SC}$.
In all $d$, generic scaling is found,  {\it i.e.\ } the flow always 
reaches one of these fixed points.
For $d<2$, $F_{\rm SC}$ is reached from any initial condition.
For $d>2$, $F_{\rm EW}$ and $F_{\rm SC}$ become locally fully attractive. More
precisely, there exists a critical 
 value of the (bare) coupling, denoted  $\hat g_c$,   that 
separates the basins of attraction of $F_{\rm SC}$ and of $F_{\rm EW}$ and for which the flow reaches
the roughening transition  fixed point $F_{\rm RT}$.
In $d=2$, $F_{\rm RT}$ coincides with $F_{\rm EW}$, 
and becomes non-Gaussian for larger dimensions. 

\begin{table}[tp]
\caption{\label{tab1} Exponent values in integer dimensions. The average numerical values are
extracted from \cite{tang92}. To our knowledge, no estimates of $\omega$ are 
available in the literature. }
\begin{ruledtabular}
\begin{tabular}{lcccc}
  $d$  &   1   &   2   &   3   &   4   \\ \hline
 $\chi$ \cite{canet11} & 0.50 & 0.33 & 0.17 & 0.075 \\
 $\chi$ (numerics) &  0.50 & 0.38 & 0.30 & 0.24 \\
$\omega$  \cite{canet11} & 0.817 & 0.70 & 0.63 & 0.54 
\end{tabular}
\end{ruledtabular}
\end{table} 

Because of Eq.\ (\ref{eqg}) and since the fixed point coupling constant 
${\hat g}^*_{\rm SC}$ is non zero, $F_{\rm SC}$ is characterized by a single exponent.
We thus only discuss below the values obtained for $\chi$ \cite{canet10}.
Table~\ref{tab1} and Fig.~\ref{fig1}(b) contain our estimates for the roughness exponent $\chi$ and  the subleading exponent $\omega$.
For $d\lesssim 4$, $\chi$ decreases almost linearly with $d$,
with the exact value $\frac{1}{2}$ (respectively  $\frac{2}{3}$) recovered in $d=1$
(respectively  $d=0$), and a reasonable but deteriorating
agreement with numerical values for $2\le d\le 4$. In higher dimensions,
$\chi$ increases with $d$, at odds with both numerical values and 
the scenario of $d=4$
being an upper critical dimension beyond which $\chi=0$
\cite{lassig97,colaiori01,fogedby05}.  The values for the critical exponents
   can be refined by computing the next order of our approximation scheme, which is work in progress.

Regarding $F_{\rm RT}$, we record negative values of $\chi$ for $2<d<5$,
which is reminiscent of perturbative results performed at fixed $d$ \cite{frey94}
but in contradiction with exact results dictating $\chi=0$ \cite{doty,wiese98}.
This is to be imputed to our breaking of the $Z_2$ symmetry manifest in  the Cole-Hopf formulation and non-linearly realized in the KPZ problem.

\section{One-dimensional scaling function}
\label{NLO}

We now consider the full \anz (\ref{anznlo}) to 
calculate the momentum- and frequency-dependent
 two-point correlation function of the one-dimensional problem and to
 extract from it the universal scaling function.  We recall that, due to the incidental time-reversal symmetry, the \anz (\ref{anznlo}) simplifies in $d=1$ to
  only one running function left: 
$f_\kappa^D = f_\kappa^\nu \equiv f_\kappa$ and $f_\kappa^\lambda\equiv 1$. We also drop from now on the vector arrows and set $\nu=D=1$ since these two coefficients can be absorbed in the action (\ref{ftkpz}) through an appropriate rescaling of the fields and of time and the change of coupling constant $\lambda \to \sqrt{g_b} = \lambda D^{1/2}/\nu^{3/2}$.   

\subsection{Flow equations}

 The flow equation for the function $f_\kappa$ can be obtained either 
from the flow of $\Gamma_\kappa^{(0,2)}$
 or from the one of $\Gamma_\kappa^{(1,1)}$ according to (\ref{anzgam2}). 
These two flows are equal in $d=1$ since the \anz  preserves all the symmetries including the time-reversal one.
 They   can be computed using  Eq.\ (\ref{dkgam2}),
where $G_\kappa$  is the  propagator matrix $G_\kappa=[\Gamma_\kappa^{(2)}+R_\kappa]^{-1}$ as defined in Eq.\ (\ref{dkgam}). Using the expressions (\ref{Rk})  and (\ref{anzgam2}) (specialized to $d=1$, {\it i.e.\ } setting $\nu_\kappa=D_\kappa$, $f_\kappa^D = f_\kappa^\nu \equiv f_\kappa$ and $f_\kappa^\lambda\equiv 1$) for the matrix elements of  $R_\kappa$ and $\Gamma_\kappa^{(2)}$, one finds for the propagator
\begin{equation}
G_\kappa(\omega,q) =\frac{1}{P_\kappa(\omega^2,q^2)}\left(\!\! \begin{array}{cc}
2 k_\kappa(\omega^2, q^2) &  Y_\kappa(\omega,q)\\
 Y^*_\kappa(\omega,q) & 0 
\end{array}\!\!\right) 
\end{equation}
where $k_\kappa(\omega^2, q^2) = f_\kappa(\omega^2, q^2)+ D_\kappa r( q^2/\kappa^2)$, $Y_\kappa(\omega,q)= i \omega+   q^2 k_\kappa(\omega^2, q^2) $ and
 $P_\kappa(\omega^2, q^2) = \omega^2 + \left( q^2 k_\kappa \right)^2$.
The calculation of the flow equation for {\it e.g.\ }$\Gamma_\kappa^{(0,2)}$, involving  matrix products and trace according to Eq.\ (\ref{dkgam2}), is then  straightforward. This yields for the flow of $f_\kappa$:
\begin{widetext}
\begin{eqnarray}
 \partial_\kappa f_\kappa(\varpi,p) &=& -\frac{1}{2} \int_{\bq}\;{\p_\kappa S_\kappa}(q) \Bigg\{
  \frac{1}{P_\kappa^2(\omega^2, q^2)} X_\kappa(\omega^2,q^2) \Gamma_\kappa^{(2,2)}(\bq,-\bq,\bp) +\frac{1}{ 2 P_\kappa^2(\omega^2, q^2)P_\kappa((\varpi+\omega)^2,(p+q)^2)}\nonumber\\
&\times& \Bigg[ 2 X_\kappa(\omega^2,q^2) Y_\kappa(\varpi+\omega,p+q)  \Gamma_\kappa^{ (1,2)}(-\bq,-\bp) \Gamma_\kappa^{(2,1)}(\bq,-\bp-\bq) - q^2 Y_\kappa^2(\omega,q) \Gamma_\kappa^{ (1,2)}(-\bp-\bq,\bp) \nonumber \\
&\times& \left(Y_\kappa(\varpi+\omega,p+q) \Gamma_\kappa^{ (1,2)}(-\bq,-\bp) +2 k_\kappa((\varpi+\omega)^2,(p+q)^2) \Gamma_\kappa^{(2,1)}(\bp+\bq,-\bq)\right) \nonumber\\
&-& \left( Y_\kappa^*(\varpi+\omega,p+q) \Gamma_\kappa^{ (1,2)}(\bq,\bp) +2 k_\kappa((\varpi+\omega)^2,(p+q)^2) \Gamma_\kappa^{(2,1)}(\bq,-\bp-\bq)\right) \nonumber\\
&\times& \left( q^2 (Y_\kappa^*)^2(\omega,q) \Gamma_\kappa^{ (1,2)}(\bp+\bq,-\bp)-2 X_\kappa(\omega^2,q^2) \Gamma_\kappa^{(2,1)}(\bp+\bq,-\bq) \right)  \Bigg]\Bigg\}
\label{flowf}
\end{eqnarray}
\end{widetext}
where $\bq\equiv(\omega,q)$, $\bp\equiv(\varpi,p)$, $X_\kappa(\omega^2,q^2)=\omega^2- q^4 k_\kappa^2(\omega ^2,q^2)$,
 $S_\kappa(q) = D_\kappa r(q^2/\kappa^2)$ and where the expressions for the three- and four-point vertex functions $\Gamma_\kappa^{(n,m)}$ are given in Appendix C.
The flow equation for $f_\kappa$ is hence an integral over the internal momentum  $ q$ and frequency $\omega$ 
and depends on the external  momentum  $ p$ and frequency $\varpi$.

As we intend to study   the fixed point properties,
we introduce dimensionless and renormalized quantities. 
 Momenta are measured in units of $\kappa$, {\it e.g.\ }$\tp = p/\kappa$, 
and frequencies in units of $D_\kappa \kappa^2$, {\it e.g.\ }$\tnu = \varpi/(D_\kappa\kappa^2)$.
We  define the dimensionless renormalized function
$\tf_\kappa= f_\kappa/D_\kappa$, as at the bare level, $f_{\kappa=\Lambda}=D=1$.
The sole running anomalous dimension $\eta_\kappa$ is defined by
$\kappa\partial_\kappa \ln D_\kappa=-\eta_\kappa$ so that $D_\kappa\sim \kappa^{-\eta^*}$ at the fixed point.
The critical exponents in $d=1$ are  then given by
 $z=2-\eta^*$  and $\chi =\eta^*$.
The absolute normalization  of $\tf_\kappa$ and $D_\kappa$ is fixed 
by setting $\tf_\kappa(0,0)=1$   for simplicity  \cite{bmw}.

The flow equation for the dimensionless function reads:
\begin{eqnarray}
\partial_s \tf_\kappa(\tnu,\tp) &\equiv& \partial_s \left[\frac{1}{D_\kappa} f_\kappa \left(\frac{\varpi}{D_\kappa\kappa^2}, \frac{p}{\kappa}\right)\right]\nonumber\\
  &=& \eta_\kappa \tf_\kappa(\tnu,\tp)+(2-\eta_\kappa) \tnu \;\p_{\tnu} \tf_\kappa(\tnu,\tp)\nonumber\\
 &+& \tp \;\p_{\tp} \tf_\kappa(\tnu,\tp) +\frac{1}{D_\kappa} \kappa \partial_\kappa f_\kappa(\varpi,p)
\label{eqf}
\end{eqnarray}
where $\partial_s=\kappa\partial_\kappa$. Once the substitutions for dimensionless quantities have been performed in (\ref{flowf}), the last term $\frac{1}{D_\kappa}\partial_s f_\kappa$ in (\ref{eqf}) is  dimensionless, and depends on the external dimensionless momentum $\tp$ and frequency  $\tnu$.

The dimensionless running coupling constant reads in one dimension
$\hat{g}_\kappa=\kappa^{-1}\lambda^2/D_\kappa^2$.
Its flow equation is  again reduced to its dimensional part due to Galilean invariance:
\begin{equation}
\partial_s \hat{g}_\kappa = \hat{g}_\kappa (2\eta_\kappa-1)\;,
\label{eqgg}
\end{equation} 
and one finds as expected that $\chi=\eta^*=\frac{1}{2}$  at any fixed point with $\hat{g}^*\neq 0$. The subleading exponent $\omega$ is calculated according to (\ref{calomega}) and we find values in agreement with the ones obtained with the simplified \anz of Section \ref{LO} (see Tables \ref{tab1} and \ref{tab2}).

\subsection{Numerical integration and error bars}

The numerical integration of Eqs.\ (\ref{eqf}) and (\ref{eqgg}) was performed
using standard techniques and the numerical  error stemming from the integration procedure was assessed by resorting to several system sizes, resolutions, {\it etc}. (see Appendix D). It turned out that the numerical error was negligible compared with the   error coming from the approximation scheme itself, which constitutes the dominant source of inaccuracy and which we now discuss.

Ultimately,  the error relative to the whole approximation scheme can be evaluated by comparing successive orders of approximation. However, when only a given order is available, one can assess the accuracy of this approximation 
 by varying the parameter 
$\alpha$ of the cutoff
function. Indeed, although physical quantities are obtained in the limit $\kappa\to 0$ (where $R_{\kappa=0}$ vanishes) from  the exact
NPRG flow equations (\ref{dkgam}) or (\ref{dkgam2})  and thus do not depend in principles on the choice
 of the particular profile of $R_\kappa$, any approximation introduces a 
spurious residual dependence on the regulator \cite{canetpms}. 
 This sensitivity to the cutoff can be investigated and exploited to estimate
 the accuracy  of the approximation implemented.   

 Here,  $\alpha$ was varied between 1 and 60,
 and all the physical quantities computed and discussed in  Section \ref{scale} exhibited a similar large plateau behavior.
 The  value given in the following (text and  Table \ref{tab2}) for a  quantity  corresponds to the central value on the plateau, and the error bars reflect the dispersion of this quantity around its plateau value when $\alpha$ varies in the range $[2,20]$.
 We emphasize that this procedure only provides an intrinsic error at a given order of approximation related to an artificial  regulator dependence. Accumulated experience in NPRG calculations seem to indicate that in general this procedure  yields a reasonable estimate of the order of magnitude of the committed error \cite{canetpms}. However, in some cases, these error bars may represent only lower bounds on the uncertainties. Definite error bars can be inferred from the comparison with the next order of approximation.

\begin{figure}[tp]
\epsfxsize=8cm
\epsfbox{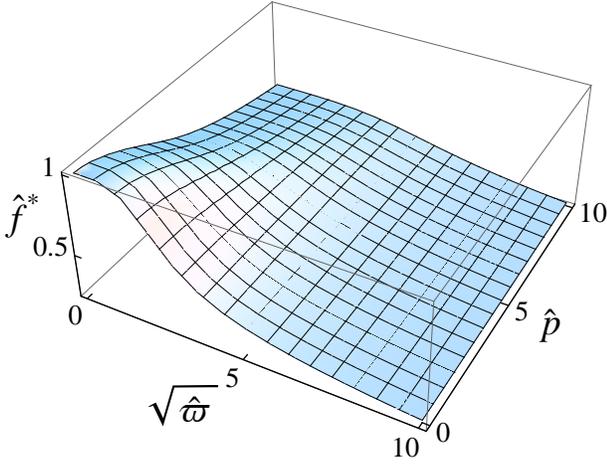}
\caption{(Color online) Typical shape of the dimensionless fixed point function $\tf^*(\tnu,\tp)$ (recorded at $s=-25$ and for $\alpha=1$).}
\label{fig2}
\end{figure}

Runs were started from various different initial conditions including 
the bare action ($\tf_\Lambda \equiv 1$) with $\hat g_\Lambda$ 
typically between 1 and  10. 
In all cases, the function $\tf_\kappa$ smoothly deformed 
from its initial shape to eventually 
 reach a fixed point where it stopped evolving,  typically after $s \lesssim-20$. In other words, the NPRG flow equations entail  generic scaling for the one-dimensional  KPZ problem, since the long time and large distance behavior is always governed by this non-trivial fixed point, yielding a scaling regime (described in Section \ref{scale}).
Fig. \ref{fig2} shows a typical shape of the fixed point function $\tf^*(\tnu,\tp)$.
It is fixed by the normalization condition to unity at vanishing 
momentum and frequency and decays as a power law when $\tnu$ and/or 
$\tp$ become large.

\subsection{Scaling function}
\label{scale}

\subsubsection{Extraction of the scaling function}

We now turn to the description of the fully attractive fixed point.
At this fixed point $\partial_s \tf_\kappa(\tnu,\tp)=0$ (in Eq.\ (\ref{eqf})) and $g^*\neq 0$ such that  $\eta_\kappa\equiv \eta^* =\frac{1}{2}$. Moreover,
 we verified numerically that the nonlinear term  $\partial_s f_\kappa/D_\kappa$ in (\ref{eqf})
 decouples (that is  $\partial_s f_\kappa/D_\kappa \to 0$) when $\tnu \gg 1$ and/or $\tp \gg 1$.
 As a consequence, at the fixed point, Eq.\ (\ref{eqf}) reduces  in the regime 
$\tnu$ and/or $\tp \gg 1$ to the homogeneous equation
\begin{equation}
\frac{1}{2} \tf^*(\tnu,\tp)+\tp \;\p_{\tp} \tf^*(\tnu,\tp)+ \frac{3}{2} \tnu \;\p_{\tnu} \tf^*(\tnu,\tp) =0.
\end{equation} 
Its general solution has the form
 \begin{equation}
 \tf^*(\tnu,\tp) = \frac{1}{\tp^{1/2}} \tzeta\left(\frac{\tnu}{\tp^{3/2}}\right)
\label{hom}
\end{equation} 
where the function $\tzeta$ cannot be determined from the homogeneous equation but  can be extracted from the  numerical solution of the full equation (\ref{eqf}) -- by
 tabulating the values $\tp^{1/2}\tf^*(\tnu,\tp)$ against the ratios $\tnu/\tp^{3/2}$.
We observe that the fixed point function $\tf^*(\tnu,\tp)$ is regular for all $\tnu$ and $\tp$ (see Fig. \ref{fig2}), which
 allows us to deduce the limits of $\tzeta$. First $\tzeta(0)$ has to be finite for the limit $\tnu\to 0$ to exist
and followingly $\tf^*\sim \tzeta(0)\tp^{-1/2}$ for $\tp \gg 1$ at fixed $\tnu$. Second, $\tzeta$ must behave as $\tzeta(x)\sim \tzeta_\infty x^{-1/3}$ as $x\to \infty$ for the limit $\tp \to 0$ to exist
 and followingly $\tf^*\sim \tzeta_\infty\tnu^{-1/3}$ for $\tnu \gg 1$ at fixed $\tp$.
 We show below that the form of the solution (\ref{hom}) entails scaling for the (dimensionful) correlation function $C(\varpi, p)$.

We now consider the fixed point dimensionful function 
\begin{equation}
f(\varpi,p) = D_\kappa \tf^*(\tnu,\tp) = D_\kappa \tf^*\left(\frac{\varpi}{D_\kappa \kappa^2},\frac{p}{\kappa}\right).
\end{equation} 
The physical limit is obtained when the running scale $\kappa$ tends to zero at fixed values of $\varpi$ and $p$.
 This limit is precisely equivalent to  $\tnu \gg 1$ and/or  $\tp \gg 1$, which corresponds to the scaling regime described above where $\tf^*$ takes the form (\ref{hom}). 
Moreover, when $\kappa\to 0$, $D_\kappa$ behaves as a power law $D_\kappa = D_0 \kappa^{-1/2}$ where $D_0$ is a non-universal constant. 
Hence, the physical dimensionful function $f$ is expressed in terms of the scaling function $\tzeta$ as
\begin{equation}
f(\varpi,p) = \frac{D_0}{p^{1/2}} \tzeta\left(\frac{1}{D_0}\frac{\varpi}{p^{3/2}}\right).
\label{fk0}
\end{equation} 
The  correlation function is defined, upon inverting the matrix $\Gamma^{(2)}$ of the two-point vertex functions, by
\begin{equation}
C(\varpi,p) = - \frac{\Gamma^{(0,2)}(\varpi,p)}{|\Gamma^{(1,1)}(\varpi,p)|^2} = \frac{2 f(\varpi,p)}{\varpi^2 + p^4 f(\varpi,p)^2}
\label{C}
\end{equation}
where the \anz  (\ref{anzgam2}), with  $f_\kappa^D = f_\kappa^\nu \equiv f$,  $f_\kappa^\lambda\equiv 1$, and $\nu=D=1$, has been used in the second equality.
Note that the definition (\ref{C}) of the correlation function corresponds in real space to the connected mean value $\langle h(t,x) h(0,0)\rangle_c$
 which differs by a factor $-2$ from the definition (\ref{defC}) in the co-moving frame. Restoring the $\nu$ and $D$ coefficients also yields an additional  factor, such that 
the correlation function defined by (\ref{C})  finally relates to the Fourier transform of the correlation function defined by   (\ref{defC}) {\it via} an overall  multiplicative factor
\be
 C_0 =  -\frac{1}{2} \frac{\nu^2}{D}. \label{C_0}
\ee

Replacing in (\ref{C}) the dimensionful fixed point function $f$ by its expression (\ref{fk0}) yields
\begin{eqnarray}
C(\varpi,p) &=&  \frac{2}{p^{7/2}}\frac{D_0 \tzeta \left(\frac{1}{D_0}\frac{\varpi}{p^{3/2}}\right)}{\varpi^2/p^3 +  D_0^2 \tzeta^2 \left(\frac{1}{D_0}\frac{\varpi}{p^{3/2}}\right)} \label{Cscal}\\
 &=&  \frac{2}{p^{7/2}} \frac{1}{D_0} \frac{ \tzeta \left(\frac{\tnu}{\tp^{3/2}}\right)}{{\tnu^2}/{\tp^3}+\tzeta^2 \left(\frac{\tnu}{\tp^{3/2}}\right)} \label{Ccalc}\\
 &\equiv&  \frac{2}{p^{7/2}} \frac{1}{D_0} \Frond\left(\frac{1}{D_0}\frac{\varpi}{p^{3/2}}\right) \label{defFrond}.
\end{eqnarray}
Eq.\  (\ref{Cscal}) shows that the correlation function takes a scaling form at  large distance and long time. 
 Let us  emphasize once more that we did not {\it assume} scaling,  the latter  emerges
  from the fixed point solution of the flow equation starting from any reasonable microscopic initial condition.
The scaling function $\Frond$ is hence universal, up to an absolute normalization and a rescaling of its argument by a multiplicative constant. 

\begin{figure}[tp]
\epsfxsize=8.5cm
\hspace{-.5cm}\epsfbox{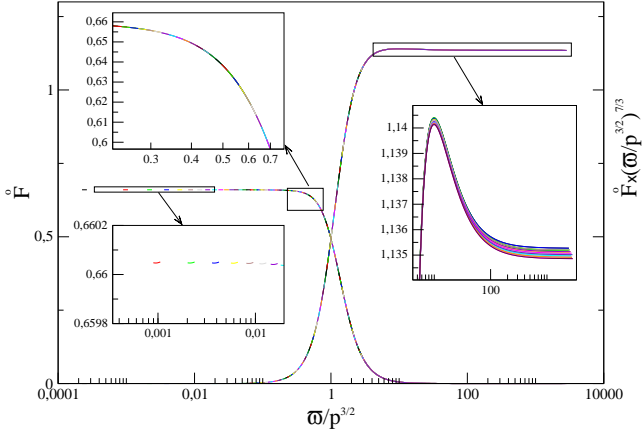}
\caption{(Color online) Leftmost  curve: Scaling function $\Frond\left(\frac{\tnu}{\tp^{3/2}}\right)$ corresponding to the data collapse (\ref{defFrond}), with the two  insets zooming different parts to show the very high quality of the collapse (note the vertical scales); the various colors  (gray shades) differentiate the contributions of distinct values of $\tnu$. Rightmost curve: Scaling function $\Frond\left(\tau = \frac{\tnu}{\tp^{7/3}}\right)$ multiplied by $\tau^{7/3}$ to illustrate the power-law behavior of the tail $\Frond \sim \tau^{-7/3}$ (see text). Note that the small dispersion visible in the inset zooming the tail for $\tau \gtrsim 100$ is hence magnified by a factor of order $100^{7/3}$.}
\label{fig3}
\end{figure}

According to 
 Eqs.\ (\ref{Ccalc}) and (\ref{defFrond}), this function $\Frond$ can be computed from   the numerical data for the dimensionless fixed function, selecting its values for  arguments in the range $\tnu$ and/or $\tp\gg 1$ which corresponds to  the scaling regime. We proceeded  for  various initial conditions and for different values of the parameter $\alpha$.
 In all cases, we  observed for each value of $\alpha$  the expected data collapse --  generating the  scaling function $\Frond$ -- with a very high precision, as illustrated  in Fig. \ref{fig3}.

\subsubsection{Normalizations}

Our aim is now to confront our scaling function $\Frond$ with the exact result obtained in Ref. \cite{spohn04}.
The first step consists of fixing the normalizations. 
Indeed, two constants are  involved in the standard definition (as recalled  below Eq.\ (\ref{defC})) of the universal scaling function $g$ from the correlation function $C$ 
$$C(\tau,L) ={\rm const.}  \tau^{2\chi/z}\,g({\rm const.}' L^{2\chi}\tau^{-2\chi/z}),$$
 that have to be fixed.
The   scaling function $g(y)$  is normalized in \cite{spohn04} in the following way:
\be
g(y) = \lim_{t\to\infty} \frac{C\left(\left(2\lambda^2 A t^2\right)^{1/3}y,t\right)}{\left(\frac{1}{2}\lambda A^2 t\right)^{2/3}} \label{normg}
\ee
with $A\equiv D/\nu$. Then  three functions $\fy$, $\fhat$ and $\frond$ are 
introduced in \cite{spohn04}, through the definitions
\begin{eqnarray}
\fy(y) &=& \frac{1}{4} g''(y)\nonumber \\
\fhat(k) &=& 2\int_0^\infty dy\, \cos(k y) \fy(y)  \label{defsp}\\
 \frond(\tau)&=& 2 \int_0^\infty dk \,\cos(k \tau) \fhat(k^{2/3})\nonumber
\end{eqnarray}
and imposing  the additional normalization condition
\be
\fhat(0) = 1. \label{normfhat}
\ee
The last function $\frond$  is proportional to $C(\omega,p) p^{7/2}$ (see \cite{spohn04}), that is  it corresponds to our function $\Frond$  reconstructed using  (\ref{defFrond})  up to normalization factors. The analog of the functions  $\fhat$ and $\fy$, that are computed in the following,  will also be denoted with  capital letters $\Fhat$ and $\Fy$ respectively.
The precise normalization  constants between $\Frond$ and $\frond$  can be established as follows. First one  deduces from Eqs.\ (\ref{normg}) and (\ref{defsp}) that:
\be
 \frond(\tau) = -\frac{p^{7/2}}{2^{5/3}\lambda^{4/3} A^{5/3}t^{7/3}}C\left(\frac{p}{\left(2\lambda^2 A t^2\right)^{1/3}},\tau\frac{p^{3/2}}{t}\right).
\ee
Then, comparing this expression with our definition (\ref{defFrond}) and taking into account the multiplicative factor
 (\ref{C_0}), we obtain the relation
\be
\frond(\tau) = 2 \sqrt{2 g^*}\sqrt{\frac{D}{\nu A}} \Frond\left(\sqrt{2 g^*}\sqrt{\frac{\nu A}{D}}\tau\right)
\ee
where we used that the bare  value $\lambda$ is related to the fixed point coupling $g^*$
through $g^* = g_b/D_0^2 = \lambda^2 D/(D_0^2\nu^{3})$.
Relating the Fourier transforms is then straightforward:
\be
\fhat(k) = 2 \frac{D}{\nu A} \Fhat\left(\frac{k}{\left(2 g^* \frac{\nu A}{D}\right)^{1/3}}\right) \equiv \frac{1}{\Fhat_n} \Fhat\left(\frac{k}{k_n}\right).
\label{normFhat}
\ee
Finally, using the  same normalization criterion  (\ref{normfhat}) as in \cite{spohn04} gives the absolute vertical normalization
 $\Fhat_n = \frac{\nu A}{2D} = \Fhat(0)$ and the absolute horizontal normalization $k_n = (4 \Fhat_n g^*)^{1/3}$.

\subsubsection{Properties of the scaling function $\fhat$}

Let us first compare the scaling functions $\fhat$ and $\Fhat$.
The function $\fhat$ is studied in detail in \cite{spohn04} and also in \cite{colaiori01,katzav04}, 
 where interesting  features are highlighted.  According to Ref. \cite{spohn04}, the function $\fhat$ first decreases to vanish at $k_0\simeq 4.36236\dots$ and then
exhibits a negative dip of coordinates ($k_d\simeq 4.79079\dots,\fhat_d \simeq -0.0012023\dots$).
 After this dip,  the function  decays to zero with a stretched
 exponential tail, over which are superimposed tiny oscillations around zero, only apparent on a logarithmic scale. A heuristic fit of this behavior for $k\gtrsim 15$ is given in \cite{spohn04}
\be
\fhat(k) \sim 10.9 k^{-9/4}\sin\left(\frac{ k^{3/2}}{2}-1.937\right)e^{-\frac{1}{2}k^{3/2}}.
\label{heuristic}
\ee

We show below that we here recover qualitatively all these features, with reasonable estimates for the different   parameters  that characterize them.
 Following (\ref{defsp}), the function $\Fhat$ is defined from $\Frond$ by the integral
\be
\Fhat(k) = \int_0^\infty \frac{d\tau}{\pi} \cos(\tau k^{3/2}) \Frond(\tau) \label{defFh}
\ee
which has to be computed numerically. The function $\Frond$ stems from the superposition of the numerous curves involved in the data collapse of Eq.\ (\ref{defFrond}). Although the data collapse is excellent,  there subsists  a high frequency and small amplitude noise due to the re-ordering of the points (apparent with a large zoom  as presented in the insets of Fig. \ref{fig3}). As we intend to investigate the properties of the tail of $\Fhat$ with high precision, we first need to eliminate this noise.
For that,  we devise an appropriate family of analytical fitting functions to  smooth our data for $\Frond(\tau)$. 
The choice of the family of fitting functions is determined as follows.

\begin{figure}[tp]
\epsfxsize=8.5cm
\hspace{-.5cm}\epsfbox{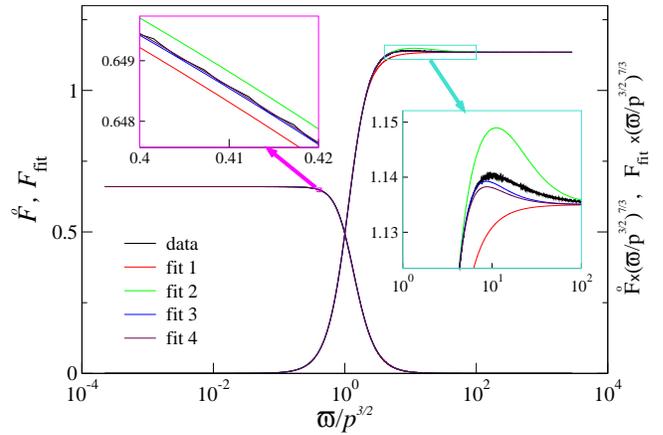}
\caption{(Color online) Illustration of the fitting procedure described in the text, using the same data for $\Frond$ as in Fig. \ref{fig3} but with unified colors (black curves: on the left $\Frond(\tau)$; on the right $\Frond(\tau)\tau^{7/3}$). The four successive orders of the fit appear superimposed in the main graph; the two insets present large zooms (note the vertical scales) to differentiate them and highlight the convergence: the higher the order, the closer to the data  (black-dotted curve) the fit lies. Note that for the right curve and  inset  $\Frond(\tau)\tau^{7/3}$ the differences are again magnified by a factor $\tau^{7/3}$.}
\label{fig4}
\end{figure}

First, they  have to reproduce the large $\tau$ behavior of $\Frond$.
  The latter can be inferred  from the limits of the function $\tzeta(x)$ established  after Eq.\ (\ref{hom}). One obtains  $\Frond(\tau)\sim \Frond_\infty \tau^{-7/3}$ as $\tau\to \infty$. 
For each value of the parameter $\alpha$, the proportionality constant $\Frond_\infty$ can be estimated graphically (up to 4-5 digits) from the curve $\tau^{7/3}\Frond(\tau)$. Moreover, the fitting functions have to be even,  and finite at the origin.  
 To satisfy these constraints, we build a family of fitting functions
 as an expansion in elementary rational polynomials of $\tau^2$ raised to the adequate power to reproduce the large $\tau$ behavior:
\begin{multline}  
\Frond_{\text{fit}}(\tau) = \left(\frac{a_{00} + a_{02}\tau^2}{1 + a_{01}\tau^2 + a_{03}\tau^4} + \frac{a_{10}}{1 + a_{11}\tau^2} \right. \\
\left.  + \frac{a_{20}}{1 + a_{21}\tau^2}+\dots \right)^{7/6}
\label{fit}
\end{multline}
with the additional constraint $ (a_{02}/a_{03} + a_{10}/a_{11} + a_{20}/a_{21}+\dots)^{7/6} =  \Frond_\infty$ \cite{NOTE}.  
 The first fit is achieved using the three independent coefficients  $a_{0i}$. The corresponding denominator introduces four non-analyticities in the points of the complex plane $\pm z_0^\pm$ with
\begin{equation}
z_0^\pm = \sqrt{\frac{1}{2 a_{03}} \left(-a_{01} \pm i\sqrt{-a_{01}^2 + 4 a_{03}}\right)}
\end{equation}
 whose coordinates were found to be very robust against the choice and the order of the fit as explained below. Note that
 these four points appeared never to lie on the imaginary axis, nor on the real axis  ($-a_{01}^2 + 4 a_{03}>0$).
We then add in turn simple monomials corresponding to purely imaginary poles up to nine independent coefficients (we observed that additional complex poles were systematically  decomposed into  two purely imaginary poles). This procedure turned out to  converge rapidly for all $\alpha$ values, as  illustrated in Fig. \ref{fig4}. In particular, the existence and the coordinates of the complex singularities $\pm z_0^\pm$ appeared to be a robust feature of the data. Moreover, they are always found to lie the closest to the real axis when other poles are included according to (\ref{fit}) \cite{except}. 
For the remaining calculations we work with the analytical expression $\Frond_{\text{fit}}$ which  reproduces faithfully our data, and hence drop in the following the index $_{\text{fit}}$.

\begin{figure}[t]
\epsfxsize=8.5cm
\hspace{-.5cm}\epsfbox{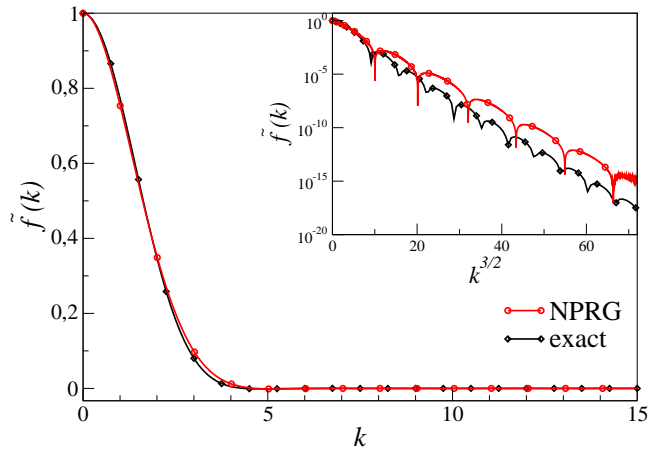}
\caption{(Color online) Comparison of the scaling function $\Fhat(k)$ (red curve with dots) obtained in this work with the exact one $\fhat(k)$ (black curve with squares) from \cite{spohn04}. The inset shows the stretched exponential behavior of the tail with the superimposed oscillations, developing on the same scale $k^{3/2}$. Note the vertical scale:  this behavior develops with amplitudes  below typically $10^{-6}$ (see text for a detailed comment of the figure).}
\label{fig5}
\end{figure}

 We computed numerically the function  $\Fhat$ from $\Frond$ according to (\ref{defFh}) and normalized it using (\ref{normFhat}), which yielded the following results. As displayed in  Fig. \ref{fig5}, the overall agreement between the NPRG scaling function $\Fhat$ and the exact one  $\fhat$  of Ref. \cite{spohn04} is excellent. We find that $\Fhat$ reproduces very accurately all the qualitative features of the exact function $\fhat$, in particular the existence of the  negative dip and the subsequent stretched exponential decay with the presence of oscillations in the tail. Indeed, we find for   the position of the first zero $k_0\simeq 4.60(6)$ and for the coordinates of the dip ($k_d \simeq 5.14(6),\Fhat_d\simeq -0.0018(6)$), which are close to the exact results, see Table \ref{tab2}.
As for the behavior of the tail of $\Fhat$, it  can be inferred analytically from
 the pole structure of $\Frond$. Let us define $\Ftilde(k)=\Fhat(k^{2/3})$, which is hence the standard Fourier transform of $\Frond$
 \begin{equation}
\Ftilde(k)=  \int_0^\infty \frac{d\tau}{\pi} \,\cos(\tau k) \Frond(\tau).
\end{equation}
As $\Frond(\tau)$ is $C^\infty$, the tail of $\Ftilde$ is dominated by the singularities of 
$\Frond(\tau)$ in the upper complex half plane lying the closest to the real axis, which are $z_0^+$ and $-z_0^-$. Denoting $z_0^+ = a_0 + i b_0$ with $b_0>0$, we obtain
$\Ftilde(k)\sim e^{- i k z_0^+} +e^{i k z_0^-}\propto e^{-b_0 k} \cos(a_0 k)$ as $k\to \infty$
 and followingly
\begin{equation}
\Fhat(k)= \Ftilde(k^{3/2})\sim  e^{-b_0 k^{3/2}} \cos(a_0 k^{3/2}) \hbox{\;\;\;\;as $k\to \infty$}.
\end{equation}
$\Fhat$ hence decays following a stretched exponential  with superimposed oscillations on the scale $k^{3/2}$ exactly  as observed in the exact solution (see (\ref{heuristic})). Regarding the tiny magnitude over which develop these features, this agreement is remarkable. Let us emphasize that within the MC approximation and the self-consistent expansion,  the stretched exponential behavior and the oscillations were also highlighted, but not on the correct scale \cite{spohn04,colaiori01,katzav04}. The NPRG  method  hence seems to provide more accurate results than  other nonperturbative approaches, and with much less inputs (no assumed scaling nor a precise  scaling form). 

A consequence is that, conversely to the MC approach for instance, the 
 coefficient of the exponential and the period of the oscillations can here be estimated -- extracted analytically
 from the values of $z_0$. 
We find $b_0\simeq 0.49(1)$ for the coefficient of the exponential, and $a_0\simeq 0.28(5)$ for the pulsation of the oscillations (checking that these values  coincide with the same quantities estimated graphically from the tail of $\Fhat$ computed numerically -- the latter being determined with much less accuracy).
These coefficients can be compared with the coefficients of the fit (\ref{heuristic})  of the tail of the exact function which are $b_0=a_0=\frac{1}{2}$.  The NPRG coefficient $b_0$  essentially matches the exact one and the pulsation $a_0$ is of the same order as the exact one, which is already highly non-trivial.  The discrepancy
 is only visible on a logarithmic scale, as illustrated in the inset of Fig. \ref{fig5}.
Let us emphasize once more that the error bars attributed here to a given quantity    reflect the weak variations of this quantity around plateau values when the $\alpha$ parameter of the cutoff function is varied. 
 It represents an error estimate intrinsic to the order of approximation under study and does not imply that the value at the next order of approximation 
 would necessarily fall within these error bars. 

\subsubsection{Scaling function in real space and universal amplitude ratio}

\begin{figure}
\epsfxsize=8.5cm
\hspace{-.5cm}\epsfbox{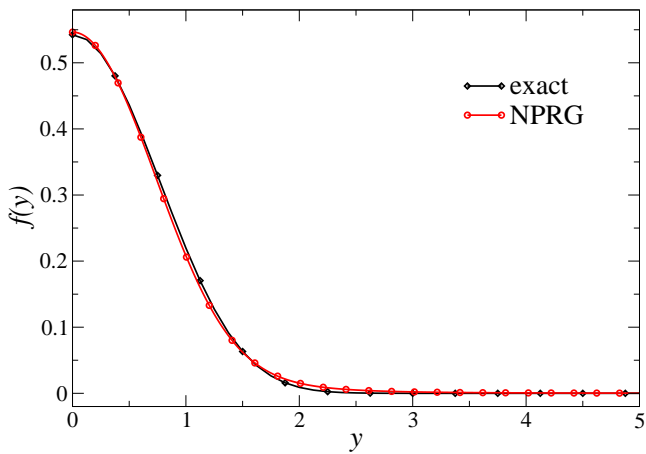}
\caption{(Color online) Comparison of the scaling function $\Fy(y)$ (red curve with dots) obtained in this work with the exact one $\fy(y)$ (black curve with squares) from \cite{spohn04}.}
\label{fig6}
\end{figure}

We finally come to the real space scaling function $\fy$ defined in (\ref{defsp}).
 We computed it numerically by Fourier transforming $\Fhat$
\begin{equation}
\Fy(y) =  \int_0^\infty\frac{dk}{\pi} \; \cos(k y )\Fhat(k).
\label{defFy}
\end{equation}
It is compared in Fig. \ref{fig6} with the exact result.
Again, the overall agreement is manifestly excellent. The NPRG function $\Fy$ reproduces  very precisely the exact one $\fy$, though it is the less accurate of our three scaling functions since it stems from the two successive numerical (oscillating) integrations (\ref{defFh}) and (\ref{defFy}) of our raw data. 
The tail is particularly sensitive to this  loss of precision. The exact function $\fy(y)$ is found to decrease as $\exp({-c y^3})$ when $y\to \infty$ in \cite{spohn04}, whereas the decay of the function $\Fy(y)$, though it starts with the correct behavior,  rapidly crosses over to a simple exponential decay $\exp({-c' y})$.

From this function can be computed the universal amplitude ratio $g_0$ with the definition given in \cite{spohn04}  
\begin{equation}
g_0 \equiv 4 \int_0^\infty dy\; y \fy(y).
\label{g0}
\end{equation}
The exact result is the Baik-Rain constant \cite{baik00} $g_0 = 1.1503944783...$ (more digits can be found in \cite{spohn04}). Here,
 the universal constant $g_0$ can be estimated by performing the  numerical integration
 corresponding to (\ref{g0}) using $\Fy$. However, this amounts to achieving three successive numerical 
 integrations from our raw function $\Frond$
 and the resulting precision is low.
Alternatively, one can compute part of the involved integrals analytically,
 using the definition given in \cite{spohn04}: $g_0\equiv g(0)$ 
 where $g(y)$ is the original scaling function proportional to the second derivative of $\fy(y)$ (see (\ref{defsp})).
Indeed, $g(y)$ can be expressed integrating twice $\fy$ through its Fourier transform $\fhat$
\begin{eqnarray}
g(y)-g(0) &=& \frac{4}{\pi}\int_0^y dv\; \int_0^v du\;\int_0^{\infty} dk\; \cos(k u)\, \fhat(k)\nonumber\\
&=&\frac{4}{\pi}\int_0^{\infty} dk\; \left(1-\cos(k y )\right)
\frac{\fhat(k)}{k^2}.
\end{eqnarray}
Taking the limit $y\to \infty$ in this expression with the change of variables $z=ky$ (and recalling that $\fhat(k=0)=1$), one deduces that in this limit:
\begin{equation}
g(y) = \frac{4}{\pi}\int_0^{\infty} dk\; \frac{1-\cos(k y )}{k^2}+\mathcal{O}(1/y),
\end{equation}
and consequently
\begin{equation}
g(0) = \lim_{y\to\infty} -\frac{4}{\pi}\int_0^{\infty} dk\; \left(1-\cos(k y )\right)
\left(\frac{\fhat(k)-1}{k^2}\right).
\end{equation}
Since the function $(\fhat(k)-1)/k^2$ is infinitely derivable, its Fourier transform vanishes in the limit $y\to \infty$  and one is left with 
 \begin{eqnarray}
g(0) &=& -\frac{4}{\pi}\int_0^{\infty} dk\; \left(\frac{\fhat(k)-1}{k^2}\right)\\
&=& \frac{6}{\pi^2}\int_0^{\infty} \frac{dk}{\sqrt{k}}\int_0^{\infty} d\tau\; \sin(\tau k^{3/2})
\tau\frond(\tau)\\
&=& \frac{2}{\pi^2}\Gamma\left(\frac{1}{3}\right) \int_0^{\infty} d\tau\;\tau^{2/3}\frond(\tau),
\end{eqnarray}
using the definitions (\ref{defsp}).
This universal quantity can thus be  determined numerically with much higher precision from $\Frond$
 by performing a unique, non-oscillating integral.
We find $g_0\simeq 1.19(1)$ which is in close agreement with the exact value.
 The defintion (\ref{g0}) suggests that the slight over-estimation is to be imputed to the contribution of the tail of the NPRG function $\Fy$, which only decays exponentially, whereas it is suppressed as  $\exp({-c y^3})$ in the exact function $\fy$.
 Note that  the MC estimate $g_0\simeq 1.1137$ \cite{colaiori01} is of  comparable accuracy with the NPRG result. However, the discrepancies  between the MCT scaling function and the exact one appear much larger (see \cite{colaiori01}), such that the  estimate of $g_0$  probably benefits from some compensation between the bulk and the tail of the MCT function. 
% (though the latter does not  provide any estimate of the error).
All our estimates for the characteristic parameters of the different scaling functions obtained in this Section are summarized in Table \ref{tab2}.
\begin{table}
%\begin{ruledtabular}
\begin{tabular}{ccc}
\hline
\hline
\;\;\;\;\;\;\;\;quantity\;\;\;\;\;\;\;\; & \;\;\;\;\;\;\;\; exact\;\;\;\;\;\;\;\; & \;\;\;\;\;\;\;\; NPRG \;\;\;\;\;\;\;\;  \\\hline
 $g_0$ &  1.15039 & 1.19(1)   \\ % 1.149(18) \\
 $k_0$ & 4.36236  &   4.60(6)   \\ %4.21(4)   \\
 $k_d$ &4.79079 & 5.14(6)  \\% 5.07(6) \\
 $\fhat_d$ & -0.00120 &  -0.0018(6)  \\% -0.013(1)\\
 $a_0$ &  $\frac{1}{2}$ & 0.28(5) \\% 0.17(1)\\
 $b_0$ &  $\frac{1}{2}$ & 0.49(1) \\% 0.34(1)\\
 $\omega$ & -- & 1.0(1)  \\ \hline \hline
\end{tabular}
%\end{ruledtabular}
\caption{Characteristic parameters of the different scaling functions, from the exact results of Ref.\  \cite{spohn04} and from this work:  i) relative to $\fy$, universal amplitude ratio $g_0$; ii) 
 relative to $\fhat$:  position of the first zero $k_0$,  coordinates of the negative dip ($k_d$,$\fhat_d$), coefficient of the stretched exponential $b_0$, pulsation of the oscillations $a_0$; iii) correction to scaling exponent $\omega$. The error bars  reflect the weak variations  around plateau values when the $\alpha$-parameter of the cutoff function is varied between 2 and 20.}
\label{tab2}
\end{table}

\section{Conclusion}

In this work, we have presented a general framework to investigate the KPZ equation using the NPRG method.
We have proposed a detailed and revisited analysis of the symmetries of the KPZ equation, both the standard 
symmetries -- Galilean and shift --  and their versions gauged in time, as well as the one-dimensional 
time-reversal symmetry and a nonlinearly realized  $Z_2$ symmetry.
We have derived general Ward identities associated with these symmetries (except for the $Z_2$ one)
and proposed a convenient geometric interpretation of the Galilean symmetry in terms of 
covariant time derivatives. 

We have then devised an approximation scheme based on an  ansatz for $\Gamma_\kappa$ which, 
in contrast to the 
standard derivative expansion often used in the NPRG framework, preserves the momentum and frequency structure of the vertex functions. 
This  ansatz,  constrained by the symmetries,
 allows one to compute correlation functions in a nonperturbative, yet systematically improvable,  way.
We explicitly  implemented the minimal order in the response field of this approximation scheme, 
 specifically developing two applications: the determination of critical exponents in physical  
dimensions (with a simplified  ansatz) and the 
 computation of  universal scaling functions in one dimension.

 We found that, without any input other than the bare action and its symmetries
(in particular without assuming scaling), 
  the renormalization
group flow is generically attracted toward a
strong-coupling fixed point -- which roots the existence of 
generic scaling --  or to the EW fixed point for $d>2$ and $\lambda<\lambda_c$, yielding for the first time 
the full correct phase diagram of the KPZ equation within a RG approach.

 The estimates of the critical exponents obtained  in dimensions two and three compare 
reasonably with results from simulations, though the accuracy of these exponents decreases with the dimension.
We are confident that these results, which were obtained using what remains, 
finally, a rather crude approximation, will get better as higher order approximations  are studied -- which is hopefully feasible in practice.
Indeed, we know from experience that within the NPRG framework, the convergence is generally very fast, such that pushing it  to even a slightly higher order  can eventually provide excellent results \cite{canetpms}.
Similarly, we are unable, at this current order of approximation, to settle on the existence 
of an upper critical dimension, but we should be able to solve this issue at next orders.

The scaling functions $\Frond$, $\Fhat$ and $\Fy$ obtained in one dimension, on the other hand, 
compare very accurately with the exact ones. 
In particular, we  recover the stretched exponential with tiny superimposed oscillations  
 of the tail of $\fhat(k)$  on the correct scale $k^{3/2}$ and with the correct coefficient $b_0\simeq 1/2$ for the exponential decay. We also obtained an accurate 
estimate of the universal amplitude ratio $g_0\simeq 1.19(1)$. 
 Quantities beyond the leading scaling regime can also be computed, which we illustrated here with the determination of the correction to scaling exponent $\omega$. 
 
A further step will consist of refining the computation of critical exponents in all dimensions 
 by improving the approximation, which is in progress.
 One could also compute the probability distributions of the $h(t,\vec{x})$ field, 
which are again  known exactly in one dimension 
but not in any other, and investigate the influence of boundary conditions.
\begin{acknowledgments}

L. Canet   thanks the Universitad de la Rep\'ublica for funding and hospitality during important 
stages of this work. The authors also wish to thank  R. Blanch for his help to optimize the numerical
 codes and parallelize them, M. Pr\"ahofer and H. Spohn for providing us with their data for the scaling 
function $\fy$ and its Fourier transform $\fhat$, and M. Moore for discussions concerning the MC results. 
N. Wschebor acknowledges the support of  PEDECIBA and ANII-FCE. The numerical parallel codes were run on the 
clusters FING and PEDECIBA (Montevideo)  and on the cluster Healthphy (CIMENT, Grenoble).

\end{acknowledgments}

\begin{widetext}

\section*{Appendix A: Time-reversal Ward identities for the three- and four-point functions}

All the Ward identities for the three- and four-point vertex functions are derived by taking functional derivatives of the identity (\ref{WardTFD}) and evaluating them at uniform and static fields.  
For the remaining three-point function, one obtains in Fourier space:
\begin{eqnarray}
2 \Gamma_\kappa^{(3,0)}(\omega_1,\vec p_1;\omega_2,\vec p_2)
&&=-\frac{\nu}{D}p_1^2\mathrm{Re}\Gamma_\kappa^{(2,1)}(\omega_2,\vec p_2;-\omega_1-\omega_2,-\vec p_1-\vec p_2)
-\frac{\nu}{D}p_2^2\mathrm{Re}\Gamma_\kappa^{(2,1)}(\omega_1,\vec p_1;-\omega_1-\omega_2,-\vec p_1-\vec p_2)\nonumber\\
&&-\frac{\nu}{D}p_3^2\mathrm{Re}\Gamma_\kappa^{(2,1)}(\omega_1,\vec p_1;\omega_2,\vec p_2)
 +\frac{1}{2}
\Big(\frac{\nu}{D}\Big)^3p_1^2p_2^2p_3^2\Gamma_\kappa^{(0,3)}(\omega_1,\vec p_1;\omega_2,\vec p_2).
\end{eqnarray}
For the four-point functions, we report the four additional independent identities obtained for uniform and static fields: 
\begin{eqnarray}
2 \mathrm{Re}\Gamma_\kappa^{(1,3)}(\omega_1,\omega_2,\omega_3)
&&=-\frac{\nu}{D}p_1^2\Gamma_\kappa^{(0,4)}(\omega_1,\omega_2,\omega_3)\nonumber\\
2 \mathrm{Im}\Gamma_\kappa^{(2,2)}(\omega_1,\omega_2,\omega_3)
&&=-\frac{\nu}{D}p_2^2\mathrm{Im}\Gamma_\kappa^{(1,3)}(\omega_1,\omega_2,\omega_3)
-\frac{\nu}{D}p_1^2\mathrm{Im}\Gamma_\kappa^{(1,3)}(\omega_2,\omega_1,\omega_3)\nonumber\\
2 \mathrm{Re}\Gamma_\kappa^{(3,1)}(\omega_4,\omega_3,\omega_2)
&&=\frac 1 2 \Big(\frac{\nu}{D}\Big)^3 p_2^2 p_3^2 p_4^2 \Gamma_\kappa^{(0,4)}(\omega_4,\omega_3,\omega_2)
-\frac{\nu}{D}p_4^2 \mathrm{Re} \Gamma_\kappa^{(2,2)}(\omega_3,\omega_2,\omega_4)
-\frac{\nu}{D}p_3^2 \mathrm{Re} \Gamma_\kappa^{(2,2)}(\omega_2,\omega_4,\omega_3)\nonumber\\
&&
-\frac{\nu}{D}p_2^2 \mathrm{Re} \Gamma_\kappa^{(2,2)}(\omega_3,\omega_4,\omega_2)\nonumber\\
\end{eqnarray}
 and finally
\begin{equation}
p_1^2\mathrm{Im}\Gamma_\kappa^{(3,1)}(\omega_2,\omega_3,-\omega_1-\omega_2-\omega_3)
+\mathrm{3\,perm.}
=-\frac 1 2 \Big(\frac{\nu}{D}\Big)^2 p_1^2 p_2^2 p_3^2 \mathrm{Im} \Gamma_\kappa^{(1,3)}(-\omega_1-\omega_2-\omega_3,\omega_1,\omega_2)
+\mathrm{3\,perm}
\end{equation}
where `perm.'  refers to the following: on the left-hand side,  the permutation of the last (implicit) frequency-momentum argument (corresponding to the $\tilde \varphi$ leg) together with the $p^2$ factor in front, from $\omega_1$ over  to  $\omega_2$, $\omega_3$ and $-\omega_1-\omega_2-\omega_3$;  on the right-hand side, the permutation of the first frequency-momentum argument (corresponding to the $\varphi$ leg) together with the complementary product of $p^2$ in front, from $-\omega_1-\omega_2-\omega_3$ over to 
 $\omega_1$,  $\omega_2$ and $\omega_3$.\\

\end{widetext}

\section*{Appendix B: Validity of the approximation scheme}

The approximation scheme presented in Section \ref{ANZ} is closely inspired by the BMW scheme, which provides an accurate description of the momentum dependence of two-point functions at equilibrium \cite{bmw}. The BMW scheme thus stands as the most adequate approximation scheme for the KPZ problem, both because we are intrinsically interested in the momentum and frequency structure of the two-point functions and  because the derivative nature of the interaction vertex seems to require one to go beyond a simple derivative expansion \cite{canet05}.
However, the implementation of the BMW scheme for the KPZ problem is hindered by the very demanding symmetries of the KPZ action. As a consequence, it cannot be directly performed  but has to be adapted regarding  three main aspects.

First, the standard BMW approximation at equilibrium consists, at leading non trivial order, of performing  truncations on the three- and four- point functions, keeping only the leading order in their internal momenta since high momentum contributions are suppressed by the cutoff function (see \cite{bmw} for a full justification of the scheme).
In the KPZ problem, as stressed in Section \ref{ANZ}, the Galilean symmetry relates  $n$-point functions with different $n$ and different momenta in a complicated way such that a naive expansion of the three- and four- point functions in their internal momenta and frequencies would spoil these relations. To overcome this difficulty, we propose here instead to devise an \anz for $\Gamma_\kappa$ that manifestly satisfies all the symmetries and such that they are hence automatically preserved in subsequent calculations (functional derivatives with respect to the fields). 
This \anz  imposes in turn the form of the three- and four- point functions. It appears that these functions retain some internal momentum and frequency dependencies that would be  neglected in a direct implementation of the BMW approximation but which are here necessary to fulfill the symmetry constraints. Hopefully, these additional subleading terms are not expected to deteriorate the quality of the  approximation.

Second, another specificity inherent to the KPZ problem resides in the absence of regulator on the frequency sector, which is prevented by the Galilean-gauged symmetry  (see Section \ref{cutcut}). As the whole BMW scheme relies on the presence of the cutoff term to perform expansions in internal  momenta  and frequencies, the absence of such a cutoff in  the frequency sector would render an expansion in internal frequencies unjustified.  Hopefully, the very same symmetry also cures this problem.
Indeed, if vertices are expanded in momenta, then the Galilean-gauged symmetry fixes the frequency dependence associated with  $\varphi$-legs (see {\it e.g.\ }Eq.\ (\ref{Gal21gauged})). Accordingly, no internal frequency expansion is needed for the $\varphi$-legs  once an expansion in internal momenta is performed. 

Finally, the previous discussion applies to the $\varphi$-legs. However, the $\tilde \varphi$-legs are treated in the proposed \anz (\ref{anznlo}) in a much cruder way since $\Gamma_\kappa$ is simply expanded at quadratic  order in that field. The quality of such an expansion has been analyzed in equilibrium statistical mechanics  both within the derivative expansion (see, {\it e.g.\  }\cite{berges02,canetpms}) and within the BMW approximation scheme  \cite{bmw2}, and  it seems to give reasonable results.  
Nevertheless, it probably constitutes the main source of inaccuracy of the present scheme. We leave for future work the improvement of the $\tilde \varphi$ dependence of the approximation.

\begin{widetext}
\section*{Appendix C: $n$-point functions at the minimal order of the approximation scheme}

To compute the three- and four- point functions at vanishing field ensuing from the \anz (\ref{anznlo}) involves  to take functional derivatives
 of the functions $f_\kappa^X(-\tilde D_t^2,-\nabla^2)$ ($X=D,\nu,\lambda)$ with respect to the field $\varphi$
and then to evaluate the result at  $\varphi=0$.
 This can be done using the expression of the series expansion (\ref{serieformelle}):
\begin{equation}
\frac{\delta f_\kappa^X(-\tilde D_t^2,-\nabla^2)}{\delta \varphi(t_1,\vec x_1)} 
=\sum_{m=1,n=0}^\infty a_{mn} \sum_{k=0}^{m-1}(-\tilde D_t^2)^k
\frac{\delta (-\tilde D_t^2)}{\delta \varphi(t_1,\vec x_1)} (-\tilde D_t^2)^{m-k-1}(-\nabla_x^2)^n
\end{equation}
with
\begin{multline}
 \frac{\delta (-\tilde D_t^2)}{\delta \varphi(t_1,\vec x_1)}
=\lambda\left(\nabla_x (\delta(t-t_1)\delta^{(d)}(\vec x-\vec x_1))\cdot \nabla_x (\partial_t-\lambda \nabla_x \varphi({\bf x}) \cdot \nabla_x) \right.\\
\left.+(\partial_t-\lambda \nabla_x \varphi({\bf x}) \cdot \nabla_x) (\nabla_x (\delta(t-t_1)\delta^{(d)}(\vec x-\vec x_1))\cdot \nabla_x)\right).
\end{multline}
When Fourier transforming,
\begin{equation}
 \frac{\delta (-\tilde D_t^2)}{\delta \varphi(t_1,\vec x_1)}\to -i \lambda \vec p_1 \cdot (\vec p_1+\vec p) (\omega_1+2\omega).
\end{equation}
One obtains for instance
\begin{equation}
\Gamma_\kappa^{(1,2)}(\omega_1,\vec p_1,\omega_2,\vec p_2) =i\lambda\sum_{m=1,n=0}^\infty a_{mn} 
\times \sum_{k=0}^{m-1}\omega_2^{2k}
\left[\vec p_1\cdot (\vec p_1+\vec p_2) (\omega_1+2\omega_2)\right] (\omega_1+\omega_2)^{2(m-k-1)}((\vec p_1+\vec p_2)^2)^n 
+ (2\leftrightarrow 3). 
\end{equation}
and thus:
\begin{eqnarray}
\Gamma_\kappa^{(1,2)}(\omega_1,\vec p_1,\omega_2,\vec p_2) =&&\frac{i\lambda}{\omega_1}
\left[ \vec p_1\cdot (\vec p_1+\vec p_2) \left( f_\kappa^D ((\omega_1+\omega_2)^2,(\vec p_1+\vec p_2)^2)-
f_\kappa^D (\omega_2^2,(\vec p_1+\vec p_2)^2)\right) \right.\nonumber\\
&&\left.-\vec p_1\cdot \vec p_2 \left( f_\kappa^D (\omega_2^2,\vec p_2^{\,2})-
f_\kappa^D ((\omega_1+\omega_2)^2,\vec p_2^{\,2})\right)\right].
\label{gam12}
\end{eqnarray}
All the other functions can be computed the same way.
The three-point functions read
\begin{eqnarray}
\Gamma_\kappa^{(3,0)}(\omega_1,\omega_2;\vec p_1,\vec p_2) &=& 0\\
\Gamma_\kappa^{(0,3)}(\omega_1,\omega_2;\vec p_1,\vec p_2) &=& 0\\
\Gamma_\kappa^{(2,1)}(\omega_1,\omega_2;\vec p_1,\vec p_2) &=& \lambda \vec p_1.\vec p_2 f_\kappa^\lambda\left((\omega_1+\omega_2)^2, (\vec p_1+\vec p_2)^2\right)
+\lambda \vec p_1.\vec p_2 \frac{\omega_1}{\omega_2}\left[f_\kappa^\lambda\left((\omega_1+\omega_2)^2, \vec p_1^{\,2}\right) - f_\kappa^\lambda\left(\omega_1^2, \vec p_1^{\,2}\right) \right] \nonumber \\
&+& \lambda \vec p_1.\vec p_2 \frac{\omega_2}{\omega_1}\left[f_\kappa^\lambda\left((\omega_1+\omega_2)^2, \vec p_2^{\,2}\right) - f_\kappa^\lambda\left(\omega_2^2, \vec p_2^{\,2}\right) \right] \nonumber \\
&-& i \lambda \frac{\nu}{2D} \left\{ \vec p_1^{\,2}   \frac{\vec p_2}{\omega_2}.(\vec p_1+\vec p_2)\left[f_\kappa^\nu\left(\left(\omega_1+\omega_2\right)^2, \left(\vec p_1+\vec p_2\right)^2\right) - f_\kappa^\nu\left(\omega_1^2, \left(\vec p_1+\vec p_2\right)^2\right)\right] \right.\nonumber \\
&+& \vec p_1^{\,2}  \frac{\vec p_2}{\omega_2}.\vec p_1\left[f_\kappa^\nu\left(\left(\omega_1+\omega_2\right)^2, \vec p_1^{\,2}\right) - f_\kappa^\nu\left(\omega_1^2,\vec p_1^{\,2}\right)\right]  +  \vec p_2^{\,2} \frac{\vec p_1}{\omega_1}.\vec p_2\left[f_\kappa^\nu\left(\left(\omega_1+\omega_2\right)^2, \vec p_2^{\,2}\right) - f_\kappa^\nu\left(\omega_2^2,\vec p_2^{\,2}\right)\right] \nonumber\\
&+& \left.\vec p_2^{\,2} \frac{\vec p_1}{\omega_1}.(\vec p_1+\vec p_2)\left[f_\kappa^\nu\left(\left(\omega_1+\omega_2\right)^2, \left(\vec p_1+\vec p_2\right)^2\right) - f_\kappa^\nu\left(\omega_2^2, \left(\vec p_1+\vec p_2\right)^2\right)\right]\right\}.
\end{eqnarray}
Note that the combinations appearing in $\Gamma_\kappa^{(2,1)}$ and   $\Gamma_\kappa^{(1,2)}$ are related.
If one  denotes $\Gamma_{\kappa,D}^{(1,2)}$ the expression (\ref{gam12}) where the $D$ index labels the function ($f_\kappa^D$) 
 appearing on the right-hand side, and similarly $\Gamma_{\kappa,\nu}^{(1,2)}$ the same expression (\ref{gam12}) where the function $f_\kappa^D$ is replaced by $f_\kappa^\nu$,   one has
\begin{eqnarray}
i \mathrm{Im} \Gamma_\kappa^{(2,1)}(\omega_1,\omega_2;\vec p_1,\vec p_2) = -\frac{\nu}{2D} \left\{
 \vec p_1^{\,2}\Gamma_{\kappa,\nu}^{(1,2)}(\omega_2,\omega_1;\vec p_2,\vec p_1) + \vec p_2^{\,2} \Gamma_{\kappa,\nu}^{(1,2)}(\omega_1,\omega_2;\vec p_1,\vec p_2) \right\}. 
\end{eqnarray} 
This type of relation is used in the following to shorten the expression of $\Gamma_\kappa^{(3,1)}$.
The four-point functions are
\begin{eqnarray}
\Gamma_\kappa^{(4,0)}(\omega_1,\omega_2,\omega_3;\vec p_1,\vec p_2,\vec p_3) &=& 0 \nonumber \\
\Gamma_\kappa^{(1,3)}(\omega_1,\omega_2,\omega_3;\vec p_1,\vec p_2,\vec p_3) &=& 0 \nonumber \\
\Gamma_\kappa^{(0,4)}(\omega_1,\omega_2,\omega_3;\vec p_1,\vec p_2,\vec p_3) &=& 0 \nonumber \\
\Gamma_\kappa^{(2,2)}(\omega_1,\omega_2,\omega_3;\vec p_1,\vec p_2,\vec p_3) &=&
 -\lambda^2 \vec p_1.(\vec p_1+\vec p_3)\vec p_2.\vec p_4\left[
\frac{f_\kappa^D\left((\omega_4^2,\vec p_4^{\,2}\right)}{(\omega_1+\omega_2)\omega_2} +\frac{f_\kappa^D\left(\omega_3^2,\vec p_4^{\,2}\right)}{(\omega_1+\omega_2)\omega_1}  -\frac{f_\kappa^D\left(\left(\omega_1+\omega_3\right)^2, \vec p_4^{\,2}\right)}{\omega_1 \omega_2}  \right] \nonumber \\
&&-\lambda^2 \vec p_2.(\vec p_2+\vec p_3)\vec p_1.\vec p_4\left[
\frac{f_\kappa^D\left((\omega_4^2,\vec p_4^{\,2}\right)}{(\omega_1+\omega_2)\omega_1} +\frac{f_\kappa^D\left(\omega_3^2,\vec p_4^{\,2}\right)}{(\omega_1+\omega_2)\omega_2}  -\frac{f_\kappa^D\left(\left(\omega_2+\omega_3\right)^2, \vec p_4^{\,2}\right)}{\omega_1 \omega_2}  \right] \nonumber \\
&&-\lambda^2 \vec p_1.(\vec p_1+\vec p_4)\vec p_2.\vec p_3\left[
\frac{f_\kappa^D\left((\omega_3^2,\vec p_3^{\,2}\right)}{(\omega_1+\omega_2)\omega_2} +\frac{f_\kappa^D\left(\omega_4^2,\vec p_3^{\,2}\right)}{(\omega_1+\omega_2)\omega_1}  -\frac{f_\kappa^D\left(\left(\omega_1+\omega_4\right)^2, \vec p_3^{\,2}\right)}{\omega_1 \omega_2}  \right] \nonumber\\
&&-\lambda^2 \vec p_2.(\vec p_2+\vec p_4)\vec p_1.\vec p_3\left[
\frac{f_\kappa^D\left((\omega_3^{\,2},\vec p_3^{\,2}\right)}{(\omega_1+\omega_2)\omega_1} +\frac{f_\kappa^D\left(\omega_4^{\,2},\vec p_3^{\,2}\right)}{(\omega_1+\omega_2)\omega_2}  -\frac{f_\kappa^D\left(\left(\omega_2+\omega_4\right)^2, \vec p_3^{\,2}\right)}{\omega_1 \omega_2}  \right] \nonumber\\
&\equiv& \Gamma_{\kappa,D}^{(2,2)}(\omega_1,\omega_2,\omega_3;\vec p_1,\vec p_2,\vec p_3)
\label{gam22}
\end{eqnarray}
where in the last line the $D$ index refers to the involved function $f_\kappa^D$.
We denote $\bar\Gamma_{\kappa,D}^{(2,2)}$ the two last lines of (\ref{gam22}) with the same convention for the meaning of the index $D$.
Finally, one finds, omitting the dependence in $\vec p_i$ (which follows the same order as $\omega_i$):
\begin{eqnarray}
\Gamma_\kappa^{(3,1)}(\omega_1,\omega_2,\omega_3)=&& -\frac{\nu}{2D}\vec p_3^{\,2} \Gamma_{\kappa,\nu}^{(2,2)}(\omega_1,\omega_2,\omega_3)  -\frac{\nu}{2D}\vec p_2^{\,2} \Gamma_{\kappa,\nu}^{(2,2)}(\omega_3,\omega_1,\omega_2)  -\frac{\nu}{2D}\vec p_1^{\,2} \Gamma_{\kappa,\nu}^{(2,2)}(\omega_2,\omega_3,\omega_1) \nonumber\\
-&& i \omega_3 \bar\Gamma_{\kappa,\lambda}^{(2,2)}(\omega_1,\omega_2,\omega_3) - i \omega_2 \bar\Gamma_{\kappa,\lambda}^{(2,2)}(\omega_3,\omega_1,\omega_2) - i \omega_1 \bar\Gamma_{\kappa,\lambda}^{(2,2)}(\omega_2,\omega_3,\omega_1)\nonumber\\
+&& i\frac{\lambda^2}{\omega_3} \vec p_3.(\vec p_1+\vec p_2) \vec p_1.\vec p_2 \left[f_\kappa^\lambda\left(\left(\omega_1+\omega_2\right)^2, \left(\vec p_1+\vec p_2\right)^2\right) -  f_\kappa^\lambda\left(\omega_4^2, \left(\vec p_1+\vec p_2\right)^2\right) \right] \nonumber\\
+&& i\frac{\lambda^2}{\omega_1} \vec p_1.(\vec p_2+\vec p_3) \vec p_2.\vec p_3 \left[f_\kappa^\lambda\left(\left(\omega_2+\omega_3\right)^2, \left(\vec p_2+\vec p_3\right)^2\right) -  f_\kappa^\lambda\left(\omega_4^2, \left(\vec p_2+\vec p_3\right)^2\right) \right]\nonumber\\
+&& i\frac{\lambda^2}{\omega_2} \vec p_2.(\vec p_1+\vec p_3) \vec p_1.\vec p_3 \left[f_\kappa^\lambda\left(\left(\omega_1+\omega_3\right)^2, \left(\vec p_1+\vec p_3\right)^2\right) -  f_\kappa^\lambda\left(\omega_4^2, \left(\vec p_1+\vec p_3\right)^2\right) \right].
\end{eqnarray}
\end{widetext}

\section*{Appendix D: Procedure for the numerical integration}

This appendix is devoted to presenting the details of the numerical 
 procedure implemented to achieve the integration of flow equations, on the example of Eqs.\  (\ref{eqf}) and (\ref{eqgg}). 
 The momentum and frequency are discretized on a $\tp\times \sqrt{\tnu}$
 mesh of spacing $\Delta \tp$ and $\Delta \sqrt{\tnu}$ and sizes
 $\tp_{\text{max}}$ and $\sqrt{\tnu_{\text{max}}}$.
 The integrals over the internal momentum $\tq$ and frequency $\tom$
 in (\ref{flowf}) are performed using Simpson's rule. 
 The $\sqrt{\tnu}$ mesh is chosen because the integrand can have rather long tails
 in $\tom$ as there is no cutoff term (analogous to $r(y)$) suppressing  the high frequencies.
 The values of $\tf$ at $\tom\pm \tnu$ which do not fall on $\sqrt{\tnu}$ mesh points are evaluated using cubic interpolations. The function $\tf$ is extended outside
 the grid (for momenta $\tp+\tq$ greater than $\tp_{\text{max}}$ and/or frequencies $\tnu+\tom$  greater than $\tnu_{\text{max}}$) using power law extrapolations.

Regarding the integration over $\tq$, as the integrand falls off exponentially due to  the  (derivative of the regulator) $\p_s S_\kappa$
 term in (\ref{flowf}), the bounds $\pm \infty$ of the integral can be safely replaced by $\pm \tp_{\text{max}}$. This is not the case for the integral over the frequency $\tom$ as the decay of the integrand
 may be slow.
  We  first compute the integral on $[-{\tnu_{\text{max}}},{\tnu_{\text{max}}}]$ and then evaluate the contribution of the integral on the boundaries $[-\infty,-{\tnu_{\text{max}}}]$ and
   $[{\tnu_{\text{max}}},\infty]$ by performing in these regions the change of variable $x = \tom/\tom_{\text{max}}$ and using the extrapolated values of $\tf$.
 The precision on the momentum and frequency integral is of order $10^{-4}$
 for the typical resolutions $\Delta \tp=\Delta \sqrt{\tnu} =  1/4$ and mesh sizes $\tnu_{\text{max}} =  15^2$ and  $\tp_{\text{max}}$ from 20 to 45 increasing with the value of $\alpha$ (20 for $\alpha=0.5$ to 45 for $\alpha=14$).
 The derivative terms $\tp \p_{\tp}$ and $\tnu \p_{\tnu}$ are computed using 5-point differences.

We use an explicit Euler time stepping with a typical time step $\Delta s =- 2.10^{-5}$
 to integrate the flow equations on the renormalization time $s$, which turns out to be  stable.
In all cases, a fixed point is reached after $s\lesssim -20$. 
The fixed point functions are recorded at $s=-25$.
%The computation time to get the correlation function  is large enough
% to require the paralellization (OMP and MPI -- depending on the architecture of the cluster used) of the code.
%The runs lasted typically 10-30 days with 60 nodes in parallel.
We studied separately the influence of the resolution
($\Delta \tp$ and $\Delta \sqrt{\tnu}$),
and of the mesh sizes ($\tp_{\text{max}}$ and $\tnu_{\text{max}}$) 
on the precision level, and checked the convergence.
 The differences (in all physical quantities) 
between resolutions and/or domain sizes
 are in all cases smaller than typically $1\%$ and these numerical errors are dominated
 by the residual variations observed when varying the 
cutoff parameter $\alpha$.


\begin{thebibliography}{10}
\bibitem{NOTE}
This is done to improve the quality of the fit, otherwise the relative weight of the tail is very small in the fitting procedure.
\bibitem{except}
The only exception to $\pm z_0^\pm$ lying the closest to the real axis comes from spurious poles of the rational fraction inside the power law in (\ref{fit}) that are also zeros of the numerator and therefore do not contribute. Those artifacts appear ponctually and do not persist from an order of the fit to the other.
\end{thebibliography}
\end{document}